\documentclass{aa}  

\usepackage{graphicx}
\usepackage{xcolor}
\usepackage{subcaption}
\usepackage{multirow}
\usepackage{comment}
\usepackage{svg}
\usepackage[colorlinks,linkcolor=blue,citecolor=blue,urlcolor=blue]{hyperref}
\usepackage{txfonts}

\begin{document}

   \title{X-ray emission maps and scaling relations in IllustrisTNG and MillenniumTNG}

   \subtitle{Differences between cluster and group regimes}

   \author{Franklin Aldás
          \inst{1,2,3}
          \and
          Facundo A. Gómez\inst{3}
          \and
          Volker Springel\inst{4}
          \and
          R\"udiger Pakmor\inst{4}
          \and
          Alex Saro\inst{1,2,5,6,7}
          \and
          Tiago Castro\inst{8,1,2, 6,7}
          }

   \institute{ INAF-Osservatorio Astronomico di Trieste, Via G. B. Tiepolo 11, I-
34143 Trieste, Italy.
\email{franklin.aldas@inaf.it}
\and IFPU, Institute for Fundamental Physics of the Universe, via Beirut 2,
34151 Trieste, Italy. \and
Departamento de Astronomía, Universidad de La Serena, Avenida Juan Cisternas 1200, La Serena, Chile.
    \and
    Max-Planck-Institut fur Astrophysik, Karl-Schwarzschild-Str 1, D-85748 Garching, Germany
    \and
    Dipartimento di Fisica, Sezione di Astronomia, Università di Trieste, Via G. B. Tiepolo 11, 34143 Trieste, Italy
    \and 
    INFN – Sezione di Trieste, Via Valerio 2, 34127 Trieste, Italy
    \and 
    ICSC – Centro Nazionale di Ricerca in High Performance Computing, Big Data e Quantum Computing, Bologna, Italy
    \and
    Department of Mathematical Physics, Institute of Physics, University of S\~ao Paulo, R. do Mat\~ao 1371, 05508-090, S\~ao Paulo, SP, Brazil
             }

   \date{Received September 15, 1996; accepted March 16, 1997}

 
  \abstract
{X-ray observations are a primary probe of the intracluster medium, widely used to infer galaxy cluster masses and scaling relations. We present and validate a pipeline to generate X-ray emission maps of galaxy groups and clusters from large cosmological simulations, and use it to study the origin of deviations from self-similarity across the group-to-cluster transition.
We apply this pipeline to the Illustris-TNG300 and MillenniumTNG simulations, constructing \mbox{X-ray} emission maps and spectra based on APEC cooling functions. For simulations that do not explicitly track individual chemical abundances, we introduce a metallicity-based prescription that accurately reproduces the full spectral emission. We derive the $L_{\mathrm{X}}$--$M_{500}$, $M_{\mathrm{gas}}$--$M_{500}$, and $T$--$M_{500}$ scaling relations over $10^{12.5} \leq M_{500} \leq 10^{15.5}\,\mathrm{M}_\odot$, compare them with observational data, and quantify the hydrostatic equilibrium and spectroscopic temperature biases through synthetic X-ray analyses.
The simulated scaling relations are in good overall agreement with observations and are best described by broken power laws with a pivot at $M_{500}=10^{13.67}\,\mathrm{M}_\odot$. At high masses, the slopes are close to self-similar expectations; at lower masses the relations steepen significantly, reflecting the growing importance of AGN feedback.
X-ray hydrostatic masses are systematically underestimated by $15\%$, independently of cluster mass. When spectroscopic effects are included, the bias becomes mass-dependent, ranging from $15\%$ at low masses to $21\%$ at high masses. The recovered X-ray luminosity, measured in the $0.15$--$1\,R_{500}$ aperture, is also mass-dependent: high-mass clusters are underestimated by $18\%$, while low-mass systems show discrepancies of up to $33\%$, driven by single-temperature spectral modelling of the gas outside the core. Despite these biases, the slope and normalization of the $M_{500}$--$L_{500}$ scaling relation are only weakly affected, because the spectroscopic pipeline shifts both mass and luminosity simultaneously, absorbing a fraction of the bias along the intrinsic scaling relation rather than perpendicular to it.}


   \maketitle
%
\section{Introduction}
\label{Sec:Introduction}
Within the $\Lambda$-CDM model, structures in the Universe grow in mass in a hierarchical way~\citep{White1978}. In this framework, small initial density fluctuations in the early Universe, amplified by gravitational instability, lead to the formation of the first bound structures, small dark-matter halos. These small halos merge and accrete additional matter over time, gradually forming larger and more massive structures. This process continues in a bottom-up fashion, forming groups and clusters of galaxies~\citep{Press1974}. During this process, baryons fall into the deep potential wells generated by dark matter and are shock-heated to temperatures of $10^{7}$–$10^{8}\,\mathrm{K}$ for halos corresponding to galaxy clusters, forming the hot, diffuse intracluster medium (ICM). Because the ICM contains most of the baryonic mass in clusters and emits in X-rays, it provides a direct window into the thermodynamical state of large-scale structures and into the physical mechanisms that regulate baryonic matter on megaparsec scales~\citep{Sarazin1986, Allen2011, Bohringer2010}.

X-ray observations play a central role in the study of galaxy clusters, as they provide direct insight into the thermodynamical state of the ICM~\citep{Biffi2012, Bohringer2012, Truong2024}. The observed X-ray luminosity, temperature, and gas density profiles trace the depth of the gravitational potential well, the baryonic content, and the physical conditions of the hot plasma permeating the cluster. The X-ray spectra are characterized by a thermal bremsstrahlung continuum together with numerous atomic emission lines, whose relative intensities and spectral shapes encode detailed information on the thermodynamic structure, kinematic state, and chemical enrichment of the gas~\citep{Mazzotta2004, McNamara2007, Allen2011}. These reconstructed density and temperature profiles form the basis of fundamental scaling relations, such as $L_{\mathrm{X}}$–$M_{\rm 500}$, $M_{\rm gas}$–$M_{\rm 500}$, and $T$–$M_{\rm 500}$, linking observable X-ray quantities to the underlying halo mass. These relations are indispensable for cosmology: they enable the conversion of cluster catalogues from X-ray surveys into halo mass functions, which in turn provide constraints on key cosmological parameters, including $\Omega_{\mathrm{m}}$, $\sigma_8$, and the dark energy equation of state~\citep{Mantz2010, Pratt2019}.

In the simplest theoretical picture, clusters are assumed to form through purely gravitational processes, leading to a self-similar model of structure formation \citep{Kaiser1986, Bohringer2012}. Under the self-similar model, physical quantities such as temperature, luminosity, gas mass, and total mass scale with each other according to simple power laws, the fundamental ones are: i) $T$-$M_{\rm tot}$ scaling relation: According to the virial theorem, the kinetic energy of the intracluster gas (related to its temperature) should be proportional to the gravitational potential energy of the cluster, which scales with mass. This relationship leads to the scaling relation $T \propto M_{\rm tot}^{2/3}$, thus implying a slope of approximately 0.66 in log-log space. ii) $M_{\rm gas}$-$M_{\rm tot}$ scaling relation: In a self-similar model, it is assumed that the gas mass fraction (the ratio of gas mass to total mass) is constant across clusters. This implies that the gas mass should scale directly with the total mass, leading to a linear relationship $M_{\rm gas} \propto M_{\rm tot}$. Consequently, the slope in log-log space is expected to be 1. iii) $L_X$-$M_{\rm tot}$ scaling relation: The X-ray luminosity depends strongly on the square of the gas density, but since X-ray emission is mainly due to the Bremsstrahlung process, it also scales with $T^{1/2}$. Assuming the gas density scales with the total mass and considering that the temperature scales as $T \propto M_{\rm tot}^{2/3}$, the X-ray luminosity is expected to scale as $L_{X, \rm BOL} \propto M_{\rm tot}^{4/3}$. This results in a slope of approximately 1.33 in log-log space. If the luminosity is measured in a restricted band, the slope of this scaling relation can change.  For example, in the  X-ray soft band the slope is $L_{X, \rm soft} \propto M_{\rm tot}$~\citep{Bohringer2012}.

Self-similarity provides a natural baseline against which deviations can be interpreted as signatures of additional physical processes. Observations, however, have long demonstrated that real clusters do not follow these relations exactly, especially in the regime of galaxy groups and low-mass clusters, where the $L_{\mathrm{X}}$--$M_{\rm tot}$ and $M_{\rm{gas}}$--$M_{\rm tot}$ relations are significantly steeper than predicted by gravity alone~\citep{Reiprich2002, Maughan2007, Lovisari2015}. 
Despite this general picture, a key open question remains: to what extent are the observed departures from self-similarity physical in origin, and to what extent are they driven or amplified by observational biases in X-ray mass and temperature measurements? Cluster masses inferred from X-ray data typically rely on the assumption of hydrostatic equilibrium and spherical symmetry~\citep{Evrad1996, Comerford2007, Eckmiller2011, Lagana2013, Valdarnini2021}. Both assumptions are only approximate, since the ICM is supported not only by thermal pressure but also by non-thermal components such as turbulence, bulk motions, and cosmic rays~\citep{Lau2009, Angelinelli2020}. 
In addition, X-ray temperatures obtained from spectral fitting are affected by spectroscopic biases in multi-phase gas, which further propagate into mass estimates~\citep{Mazzotta2004, Vikhlinin2006}. Disentangling the relative impact of physical baryonic processes and measurement-induced biases is therefore essential for a correct interpretation of cluster scaling relations, particularly in the transition region between galaxy groups and massive clusters.

In this work, we present a fast and accurate pipeline to generate mock X-ray emission maps and spectra for galaxy groups and clusters in large cosmological simulations. We apply this framework to two complementary datasets: the IllustrisTNG300 simulation, which provides high mass resolution and detailed chemical abundance tracking, and the MillenniumTNG simulation, which offers an unprecedented statistical sample of massive clusters thanks to its large cosmological volume. Together, these simulations allow us to study X-ray scaling relations over more than three orders of magnitude in mass, from $M_{\rm 500} \sim 10^{12.5}\,\text{M}_\odot$ to $M_{\rm 500} \sim 10^{15.5}\,\text{M}_\odot$, spanning the full transition from galaxy groups to the most massive clusters.

A closely related study was carried out by \citet{Pop2022} and \citet{Pop2022b}, who investigated the X-ray and Sunyaev--Zel'dovich scaling relations and the hydrostatic mass bias of groups and clusters in TNG300 at $z=0$. The present work builds on and substantially extends that effort, both methodologically and in simulation statistics: we incorporate the much larger MillenniumTNG volume to reach the most massive cluster regime, develop and validate a metallicity-based spectral prescription for simulations that track only total metallicity, extend the analysis to the redshift evolution of the $L_X$--$M_{500}$ relation out to $z=1$, and quantify the spectroscopic luminosity bias together with its covariance with the mass bias.

The main goals of this paper are therefore: First, to introduce and validate an efficient and flexible pipeline for generating X-ray emission maps from large cosmological simulations. Second, to establish robust X-ray scaling relations over an unprecedented mass range by combining the statistical power of MillenniumTNG with the detailed physics of IllustrisTNG. Third, to test whether the observed departure from self-similarity in galaxy groups is physical in origin and driven by baryonic feedback processes, or rather an artifact of X-ray mass reconstruction.

This paper is structured as follows. Section~\ref{sec:Simulations} describes the simulations used in this work. Section~\ref{sec:Mocks} presents the pipeline adopted to generate the mock X-ray emission maps. Section~\ref{sec:Scaling} derives the scaling relations between X-ray luminosity, temperature, gas mass, and total halo mass. Section~\ref{Sec:Mass_baias} describes the reconstruction of the mass--luminosity scaling relation using methods analogous to those commonly employed in observational X-ray analyses, with the central regions of galaxy clusters excluded. Section~\ref{Sec:Redshift} examines the redshift evolution of the X-ray luminosity--total halo mass relation. Section~\ref{Sec:Conclusions} summarises the main results and discusses their implications. Appendix~\ref{Sec:Mass_baias_todo} presents the mass--luminosity scaling relation derived when the central regions are retained. Appendix~\ref{Ap:Metals_metallicity} demonstrates the equivalence of the pipeline when using either individual elemental abundances or the total gas-cell metallicity. Appendix~\ref{Ap:profiles} describes the parametric models adopted to fit the density and temperature profiles of galaxy clusters. 

\section{Simulations}
\label{sec:Simulations}
To investigate the formation and evolution of cosmic structures, hydrodynamical cosmological simulations are required to address the non-linear interplay of diverse physical components. These models account for the collisionless dynamics of dark matter, which provides the gravitational framework for structural assembly, alongside the background evolution driven by dark energy. Simultaneously, they treat the hydrodynamics of baryonic matter. While baryons represent a minor component of the mass budget, their evolution is dominated by non-linear physical processes, ranging from gas cooling and star formation to feedback from supernovae and active galactic nuclei that are critical in the observed properties of the structures in the Universe~\citep{Vogelsberger2020}.
  
In this work, we employ two high-volume, full-physics cosmological simulations to investigate the X-ray emission characteristics of galaxy groups and clusters:  IllustrisTNG-300 (hereafter TNG300; \citealp{Marinacci2018}) and MillenniumTNG (hereafter MTNG; \citealp{Pakmor2023}). Both simulation projects utilized the AREPO code, which solves the hydrodynamical equations using a second-order, finite-volume method on a quasi-Lagrangian moving mesh \citep{Springel2010, Weinberger2020}.

TNG300 is part of the Illustris TNG simulation suite, which is a set of fully hydrodynamical cosmological simulations with different box sizes and particle mass resolutions~\citep{Marinacci2018, Naiman2018, Nelson2018, Pillepich2018, Springel2018}. The suite includes three primary simulations: TNG-50, TNG-100, and TNG-300, each having a box size of approximately 50, 100 and 300 Mpc, respectively. Smaller boxes have better mass resolution but poorer sampling of massive structures. In fact, TNG50 only has 23 halos with masses larger than $10^{13}$ M$_\odot$, while we find 168 in TNG100 and 3545 in TNG300. As the main focus of this work is on the properties of massive objects,  we use the TNG300 simulation. 

The baryonic mass resolution in TNG300 is $1.1\times 10^{7} \text{M}_{\odot}$, with a softening length of 370 pc, while the resolution for the DM particles is $5.9\times 10^{7} \text{M}_{\odot}$, with a softening length of 1480~pc \citep{Marinacci2018, Nelson2019}. The TNG300 model can reproduce the distribution of several observed properties of galaxies such as the bimodal colour distribution~\citep{Nelson2018}, the galaxy and halo mass functions~\citep{Pillepich2018}, the evolution of abundances of chemical elements in the intra-cluster medium~\citep{Vogelsberger2018}, the fraction of quenched galaxies in different environments up to redshift $z\sim 2$~\citep{Donnari2019, Donnari2021}, and the excess of blue galaxies in disturbed galaxy groups and clusters in comparison with relaxed ones~\citep{Aldas2025, Aldas2023}, among other properties consistent with observational results.

The MTNG project represents a novel suite of simulations comprising a flagship hydrodynamical run alongside a series of dark-matter-only (DMO) realizations, which utilize consistent box volumes while varying particle counts to assess numerical convergence \citep{Hernandez2023, Pakmor2023}. The suite has been specifically designed to probe baryonic physics on Gpc scales, facilitating the study of galaxy clusters and the large-scale cosmic web~\citep{Pakmor2023}. In this work, we utilize the primary hydrodynamical simulation of the project, which evolves a periodic volume of $500\,h^{-1}\text{Mpc}$ on a side. The initial conditions employ $4320^3$ dark matter particles and an equal number of gas cells, with an adaptive softening length for gas cells, restricted to a minimum value of $\epsilon_{\rm gas} = 370\,{\rm pc}$, and a fixed softening length of  $\epsilon_{\rm DM} = 3.7\,{\rm  kpc}$ for DM particles. MTNG has a dark matter particle mass of $M_{\rm{DM}} = 1.7 \times 10^{8}\,\text{M}_{\odot}$ and an initial gas cell mass of $M_{\rm gas} = 3.1 \times 10^{7}\,\text{M}_{\odot}$ \citep{Hernandez2023, Pakmor2023}.
\begin{figure}
    \centering
    \includegraphics[trim={0cm 0cm 0.0cm 0.0cm}, clip, width=\linewidth]{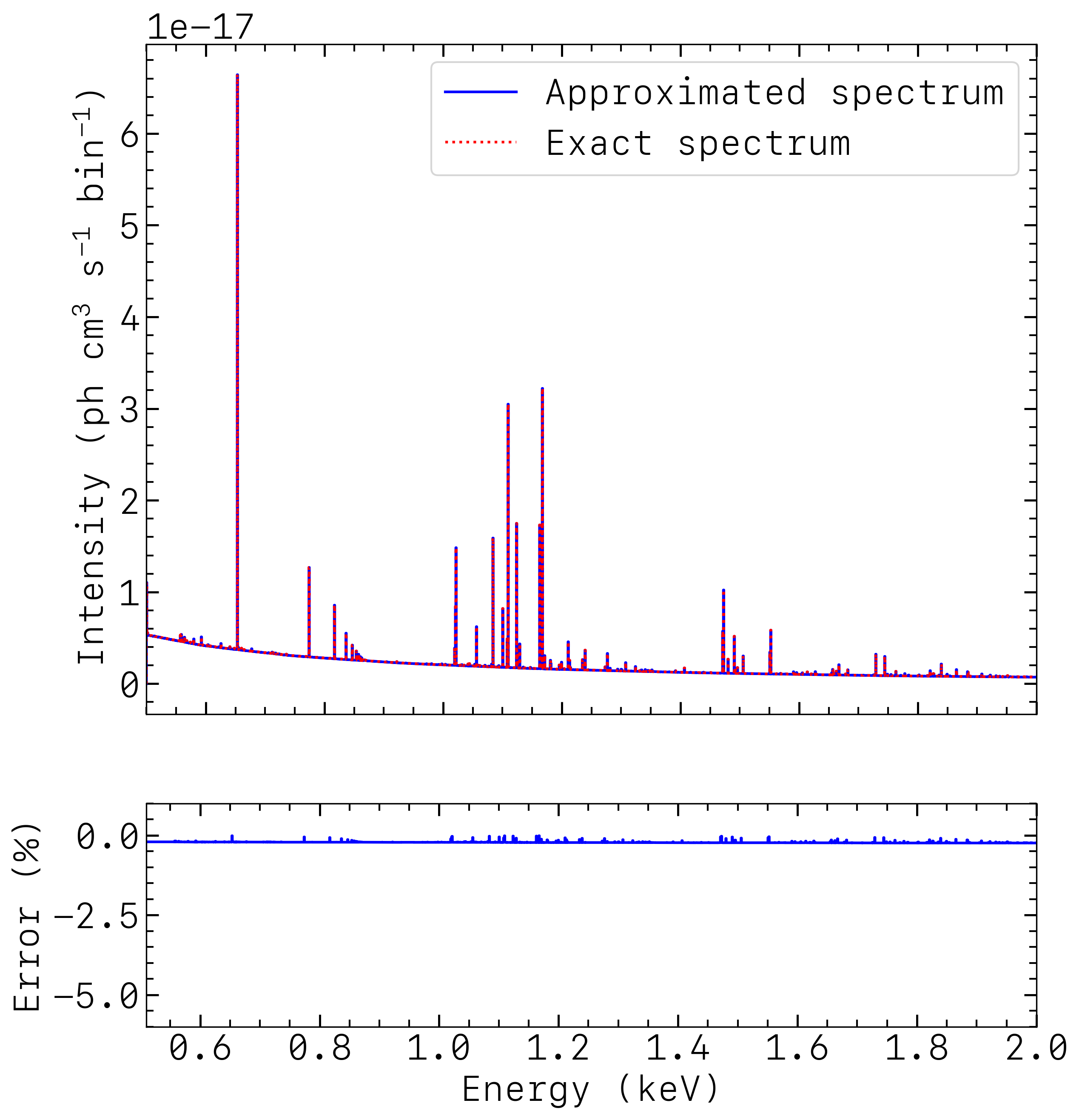}
    \caption{Comparison of X-ray spectra at $\log(T) = 7.80\,\mathrm{K}$. The spectrum computed using \texttt{pyatomdb} (blue) is compared with that obtained as a linear combination of individual elemental spectra weighted by their abundances (red). At this temperature, which coincides exactly with a grid point, the two spectra are in excellent agreement, validating the spectral interpolation scheme.}
    \label{Fig:Spectrum72}
\end{figure}

MTNG adopts essentially the same cosmological parameters and galaxy formation physics as IllustrisTNG, with the exception that it does not include magnetic fields and its tracking of chemical elements has been simplified, both done in order to save memory. This model difference leads to slightly lower black hole masses and, consequently, to higher stellar masses in massive galaxies. The MTNG volume is 15 times larger than that of TNG300; this increased statistical power provides a sample of 9 clusters with $M > 10^{15}\,\text{M}_{\odot}$, approximately 1,500 clusters with $M > 10^{14}\,\text{M}_{\odot}$, and 38,000 groups/clusters with $M > 10^{13}\,\text{M}_{\odot}$ at $z=0$, all while maintaining a mass resolution comparable to TNG300. The synergy between TNG300 and MTNG enables a robust analysis of scaling relations and X-ray emission across a broad mass range ($10^{12.5}$--$10^{15.5}\,\text{M}_{\odot}$) and X-ray luminosity span ($10^{40}$--$10^{45}\,\text{erg\,s}^{-1}$). Consistent with both TNG300 and MTNG, our analysis adopts a \citet{Planck2015} cosmology: $\Omega_{\text{m}}=0.3089$, $\Omega_{\text{b}}=0.0486$, $\Omega_{\Lambda}=0.6911$, and $h=0.6774$.

\section{Cluster X-ray emission maps}
\label{sec:Mocks}

\begin{figure}
    \centering
    \includegraphics[trim={0.0cm 0.0cm 0.0cm 0.0cm}, clip, width=\linewidth]{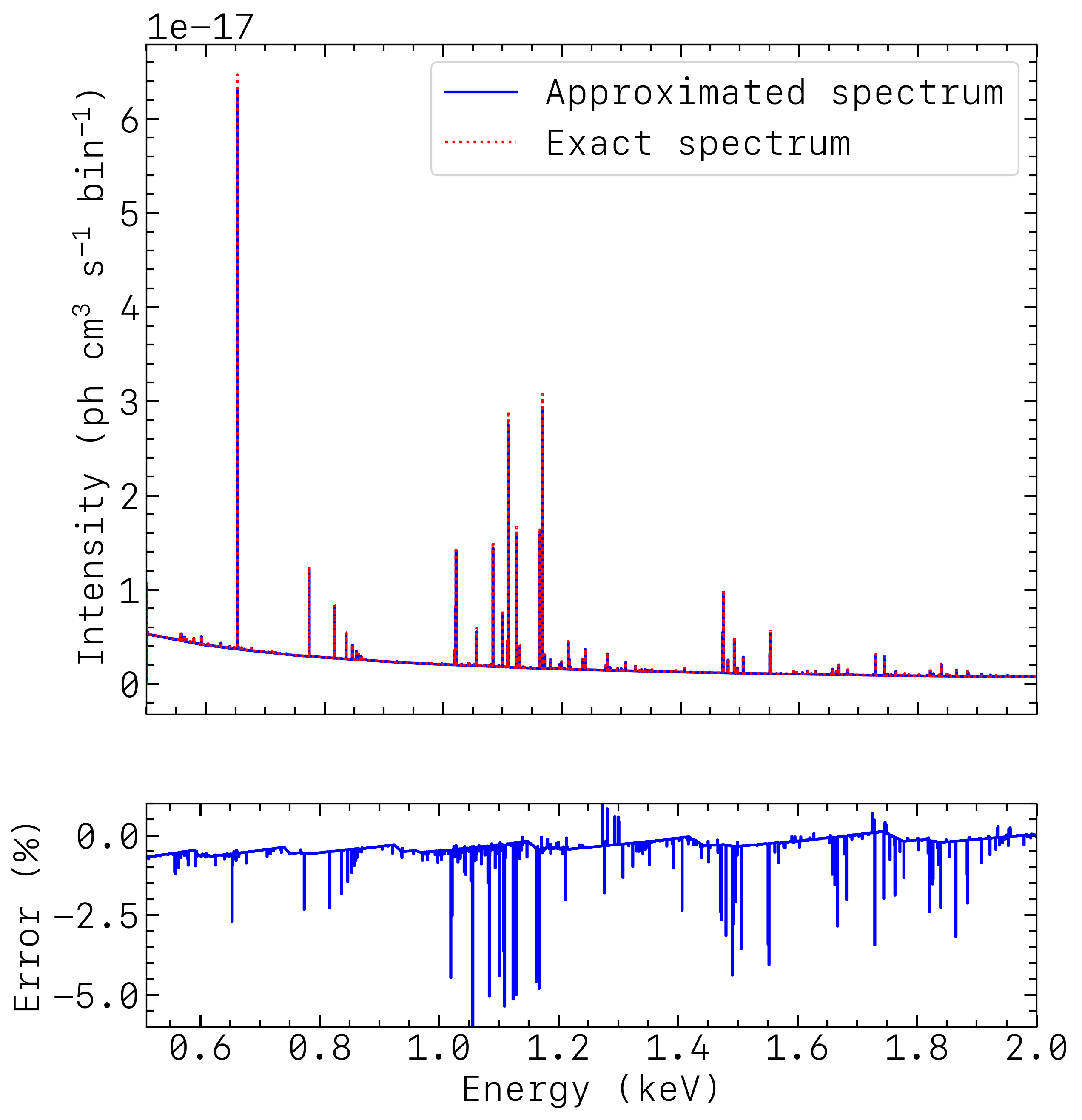}
    \caption{Comparison of X-ray spectra at $\log(T) = 7.81\,\mathrm{K}$, corresponding to the maximum interpolation distance in the temperature grid adopted in this work. The upper panel shows the spectrum computed using \texttt{pyatomdb} (blue) alongside that obtained from the weighted linear combination of elemental spectra (red). The lower panel displays the residuals between the two spectra. Minor differences are visible in the intensities of individual emission lines, while the overall spectral shape remains well reproduced. The maximum error in the integrated cooling function is approximately $3\%$.}
    \label{Fig:Spectrum73}
\end{figure}

\begin{figure*}
    \centering
    \includegraphics[width=0.9\linewidth, trim={0cm, 0.0cm, 0.0cm, 0.0cm}, clip]{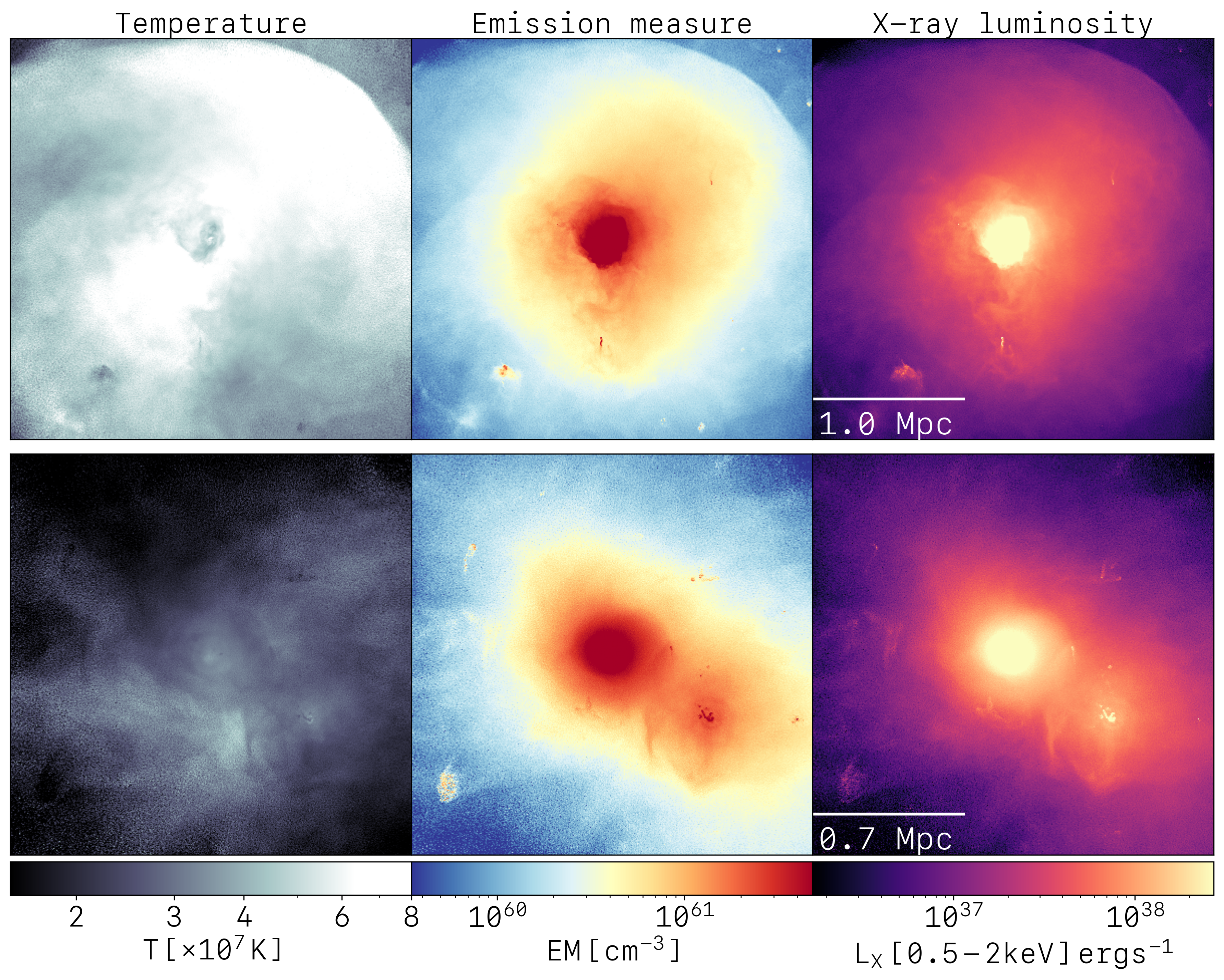}
    \caption{Maps of mass-weighted temperature (left), emission measure (centre), and soft-band X-ray luminosity in the $0.5$-$2\,\mathrm{keV}$ range (right) for two representative clusters from TNG300. The upper row shows a relaxed cluster with $M_{500} = 7.03 \times 10^{14}\,\mathrm{M}_\odot$, while the lower row shows a morphologically disturbed system undergoing a merger, with $M_{500} = 2.11 \times 10^{14}\,\mathrm{M}_\odot$. Each image is centred on the most bound particle of the halo and covers a projected area of $2 \times R_{500}$ with a line-of-sight integration depth of $6 \times R_{500}$. The relaxed cluster displays a centrally concentrated temperature and emission measure distribution, characteristic of a cool-core system, whereas the disturbed cluster shows pronounced asymmetry, shock features, and a perturbed core, reflecting the ongoing gravitational interaction.}
    \label{Fig:Properties}
\end{figure*}

Observations of galaxy clusters reveal extended X-ray emission, with bolometric luminosities spanning from \(10^{42}\) to \(10^{45}\)~erg~s\(^{-1}\). This emission originates from the hot gas permeating the intra\-cluster medium (ICM), which constitutes the dominant baryonic component of galaxy clusters. The ICM accounts for approximately 10--15\% of the total cluster mass and contains the majority of the cluster’s baryons~\citep{Allen2011}. Driven by the cluster’s gravitational potential, the gas is compressed and heated through mechanisms such as adiabatic compression and shock heating, reaching temperatures of \(10^{7}\)--\(10^{8}\)~K. At these temperatures, the ICM is almost fully ionised~\citep{Shimizu2003}. In this section, we summarise the methodology employed to generate mock X-ray emission for galaxy clusters, based on the physical properties of gas particles for the MTNG and TNG300 simulations.

For generating the simulated X-ray emission maps, we assume that the ICM is well described by the optically thin plasma approximation~\citep{Mewe1998}. Under this assumption, the emergent X-ray spectrum from the hot gas is a sum of thermal bremsstrahlung and line emission from highly ionised elements~\citep{Gilfanov1987}. We generate X-ray emission maps for galaxy clusters in \textsc{TNG300} with masses \(M_{\rm 500} \ge 10^{12.5}\,\text{M}_{\odot}\), and for clusters in \textsc{MTNG} with \(M_{\rm 500} \ge 10^{13.5}\,\text{M}_{\odot}\), following the pipeline proposed by~\citet{Barnes2021}. Here, \(M_{\rm 500}\) denotes the mass enclosed within a sphere whose mean density is 500 times the critical density of the Universe. The procedure consists of three main steps: (i) computation of the cooling function \(\Lambda(T, Z)\), which quantitatively links the thermodynamical and chemical properties of the ICM to its X-ray emissivity; (ii)~derivation of the emission measure; and (iii) calculation of the X-ray luminosity. Each of these steps is described in detail below.

i) \textit{The cooling function} $\Lambda (T, Z)$  represents the rate at which the ICM radiates energy. This function, under the assumption of collisional ionization, depends on  temperature and chemical abundance patterns. Therefore, since each gas cell has its own temperature and chemical abundance, it is necessary to simulate the X-ray emission spectrum for each gas cell. This spectrum can be generated using APEC, a code designed to model UV and X-ray emission for collisionally excited plasmas~\citep{Brickhouse2005}. The model can be accessed using the Python package \texttt{pyatomdb}. However, since most massive galaxy clusters in TNG300 and MTNG are resolved with more than \(10^7\) gas particles, generating a spectrum for each particle using \texttt{pyatomdb} is not viable in terms of computational cost. To address this, we generate spectra for each chemical element traced in TNG300 at predefined temperature values, specifically ranging from \(10^5\) to \(10^9\) K in increments of \(10^{0.02}\) K. The final spectrum is then computed as a linear combination of these elemental spectra, weighted by their abundances, and using the  spectra whose temperatures are closest to the gas cell's temperature.

The temperature of each gas cell is calculated using the internal energy $u$ and electron abundance $(x_e= n_e/n_H)$ contained in the simulation outputs, based on the following equations: 
\[
    T=(\gamma -1) u/k_B \times \mu, 
\]
and,
\[
    \mu= \frac{4}{1+3X_H+4X_H x_e}\times m_p, 
\]
where $u$ is the internal energy per unit mass of the particle, $\mu$ is the mean molecular weight, $k_B $ is the Boltzmann constant, and $m_p$ is the mass of the proton. Here, we used $\gamma = 5/3$ and $X_H=0.76$ for the adiabatic index and the hydrogen mass fraction, respectively.

To illustrate the effectiveness of the spectrum generation method, we created two spectra: one at a temperature that directly matches a grid value,  $\log (T) = 7.80 $ (Figure~\ref{Fig:Spectrum72}), and another at a temperature halfway between two grid points, $\log (T) = 7.81$, which corresponds to the maximum distance in our temperature approximation (Figure~\ref{Fig:Spectrum73}). In each case, we compared the spectrum generated using this pipeline with the spectrum generated directly with \texttt{pyatomdb}. We can see that the spectrum generated with the exact temperature (Figure~\ref{Fig:Spectrum72}) matches that derived using \texttt{pyatomdb}, while the spectrum generated with the maximum possible temperature distance (Figure~\ref{Fig:Spectrum73}) shows minor differences in emission line intensities but the overall shape of the spectra is still very similar. The maximum error of the integrated value of the cooling function is ~3\%. The abundances used for those examples correspond to  solar abundances.

 To obtain the cooling function, we also need to determine the element abundances for each gas cell. These abundances are expressed in units of solar abundances from~\citet[][AG89]{AG89} and are normalised to the number of hydrogen atoms in the cell.  In TNG300, the abundances of the relevant elements are directly traced by the simulation, requiring only a conversion to the previously detailed scale. However, MTNG tracks only total metallicity, not individual chemical elements. To estimate the abundance of each chemical element, we use the following method: for each gas cell, metal abundances are approximated by multiplying the solar abundances from AG89 scaled by a factor given by \( \theta = Z/ Z_{\odot} \), where Z is the gas cell's metallicity, and Z$_{\odot}$ is the solar metallicity from AG89. The abundance of helium is assumed to remain approximately constant, with all gas cells having a helium abundance of 0.274~\citep{Asplund2009}. The hydrogen abundance is then calculated by assuming that the remaining mass in the gas cell, after accounting for the traced metals and helium, is hydrogen. This method was validated with TNG300 by first computing galaxy cluster X-ray luminosities using the abundances of traced elements and then estimating them using only the metallicity available from the simulation outputs. The comparison revealed that the final luminosity values are closely aligned, with a mean error of  $<2\%$. See Section~\ref{Ap:Metals_metallicity} for a more in depth discussion. 

ii) \textit{The emission measure} (EM): This quantity represents the number of gas particles that emit radiation in each gas cell.  It is defined as $\text{EM}= n_e n_H \text{dV}$, where $n_e$ and $n_H$ are the electron and proton number densities, respectively, i.e. the EM is proportional to the square of the gas density. The proton density is calculated as follows:
\[
    n_H= X_H\frac{\rho}{m_p}. 
\]
Here, $\rho$ is the total density of the gas cell, $m_p$ is the proton mass ($m_p=1.67\times 10^{-25} ~ g$), and $X_H$ is the hydrogen mass fraction. The electron number density $n_e$ is calculated as $n_e= x_e\, n_H$, where the electron abundance $x_e$ is obtained directly from the simulation outputs. 

iii) \textit{X-ray luminosity} ($L_{X}$): Finally, to obtain the X-ray luminosity, we proceed as follows. The luminosity is the total energy emitted by an object per unit of time. As the medium is optically thin, the emitted luminosity can be computed as the product of the emission measure and the cooling function, integrated over the total volume and within a given energy range:
\[
    L_{X}= \int_V\int_{E_{1}}^{E_2} n_e n_H ~\Lambda(T,Z) \,{\rm d}E \, {\rm d}V.
\]
To compute the luminosity, the photon-energy integration is performed numerically by integrating the model spectrum over the selected energy range before summing over gas cells. The luminosity is evaluated using only gas cells with temperatures \(T > 10^{5}\,\mathrm{K}\) and zero star-formation rate. We exclude star-forming gas because, in the simulations, it is treated through an effective subgrid model for the multiphase interstellar medium, such that its thermal properties are not directly representative of the diffuse hot plasma responsible for the X-ray emission. We also exclude cold gas cells, as they are physically more complex, are not expected to contribute significantly to the X-ray signal in the energy range considered here, and may therefore bias the inferred luminosities. Only a small fraction of gas cells fails to satisfy these criteria, amounting to less than \(2\%\) of the total number of cells.

To illustrate the described pipeline for generating mock X-ray observations, Figure~\ref{Fig:Properties} shows two rows of images, each displaying three different properties of the galaxy clusters: temperature, emission measure, and the resulting X-ray luminosity. The first row shows a relaxed halo, while the second row shows two interacting clusters. Each figure is centred on the cluster's most bound particle, and it extends to one $R_{\rm 500}$. The left panel shows the logarithmic temperature distribution of the ICM, with colours ranging from dark purple to yellow, corresponding to cooler and hotter regions, respectively. The relaxed halo displays a well-defined rounded temperature map dominated mainly by hot gas inside $R_{\rm 500}$, with the presence of cool spots corresponding to infalling small galaxies and a cool core in the very central region of the cluster, indicating a stable configuration~\citep{Fabian1994,McDonald2019}. In contrast, the interacting halo exhibits a much less clearly defined shape, where even the cluster core has been strongly perturbed by the major merger~\citep{Santos2008}. 

The central panel presents the logarithmic emission measure, which varies as the square of the ICM density, with brighter regions indicating a higher emission measure. The relaxed halo shows a strong central concentration, indicative of a settled and dense core, while the disturbed halo has a less centrally peaked emission measure, reflecting a more spread-out and less dense core. The right panel displays the resulting logarithmic \mbox{X-ray} luminosity, integrated within the $0.5 - 2$ keV energy band, i.e.~the soft X-ray emission. The relaxed halo features a centrally bright region, signifying intense X-ray emission from the dense and well-defined core, which is characteristic of a stable cluster~\citep{Wang2023}. In contrast, the disturbed cluster exhibits noticeable asymmetry and more filamentary structures in the X-ray emission. These characteristics are the result of recent gravitational interaction during the ongoing merger, where the infall of subclusters or galaxies perturbs the ICM, leading to shock fronts, turbulence, and uneven heating of the gas.

\section{Scaling Relations}
\label{sec:Scaling}
 \begin{figure}
\centering
\includegraphics[width=\linewidth]{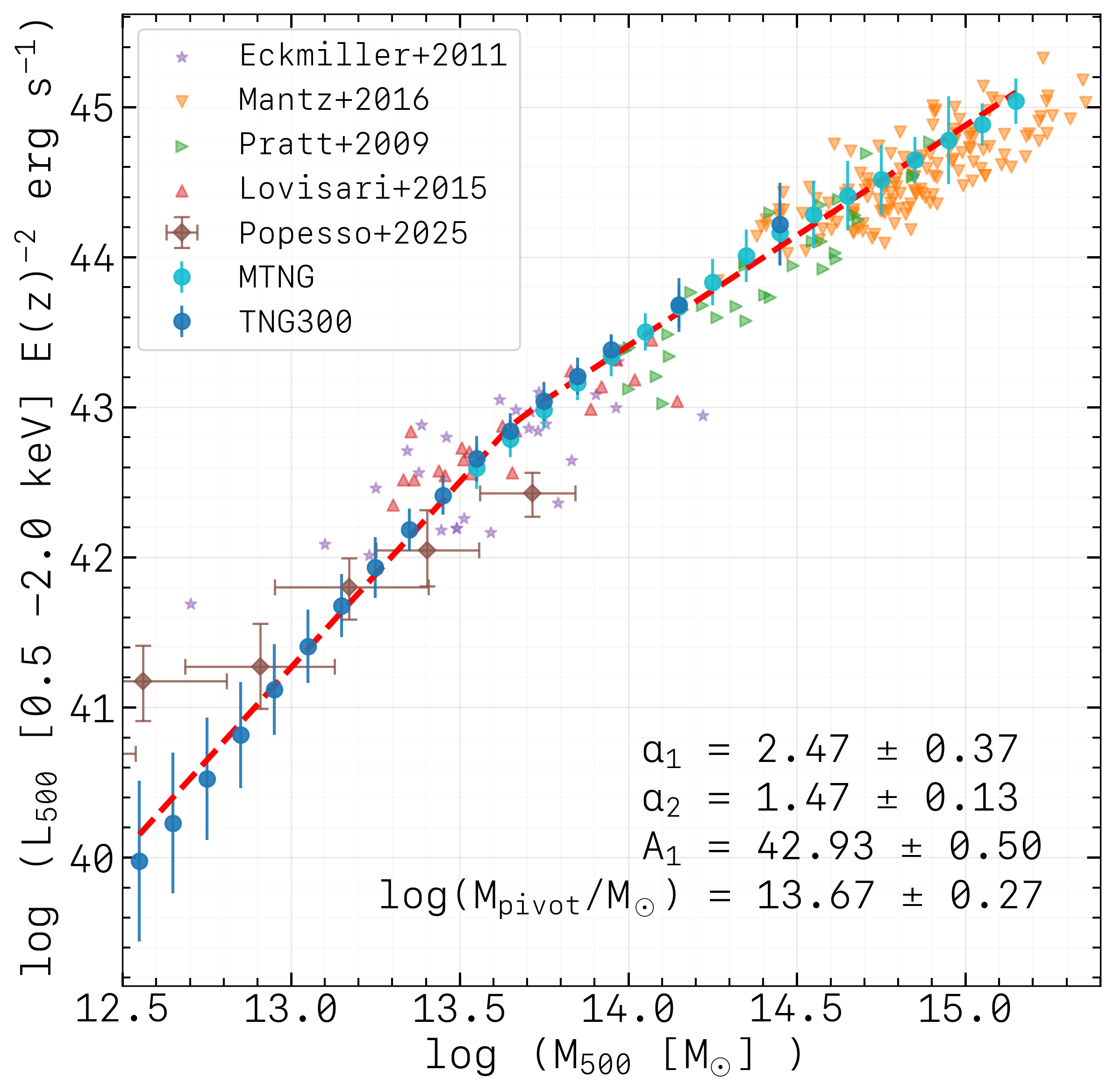}
\caption{Soft-band ($0.5$--$2.0\,\mathrm{keV}$) X-ray luminosity $L_{500}$ as a function of halo mass $M_{500}$ for the TNG300 (blue) and MTNG (cyan) simulations, compared with observational data from the literature. Data points represent the geometric mean luminosity in logarithmically spaced mass bins, and error bars denote the geometric standard deviation. The MTNG simulation extends the dynamic range to masses $M_{500} = 10^{15.5}\,\mathrm{M}_\odot$, where the statistical sample is largest. The solid lines show the best-fit broken power-law model, with the fitted pivot mass indicated. 
The simulated relations are in good overall agreement with the observational data across the full sampled mass range.}
\label{fig:M500_L}
\end{figure}
Accurate cluster mass determination is crucial for various cosmological tests, including halo counts and clustering, baryon fraction analysis, and the determination of cluster distances~\citep{Allen2011}. One common and straightforward method to estimate galaxy cluster mass is through scaling relations based on observable quantities such as the total X-ray luminosity, $L_X$~\citep{Lovisari2015, Pratt2009}. Indeed, one of cosmology's most significant scaling laws is the relationship between X-ray luminosity and $M_{\rm 500}$. This scaling relation indicates that more massive clusters host a more luminous ICM. This relation is fundamental because it links the theoretical predictions of the halo mass function to the most fundamental observable in X-ray surveys~\citep{Mantz2010b, Mazzotta2004, Pratt2019}. Establishing this relation is hence important for the derivation of cosmological parameters. For example,~\citet{Mantz2010b} constrained simultaneously the  matter density ($\Omega _m =0.23\pm 0.04$), matter fluctuation amplitude ($\sigma_8= 0.82\pm 0.05$), and the constant dark energy equation of state ($w=-1.01\pm0.20$). 

In this work, using TNG300 and MTNG, we are able to cover a large range of cluster masses with good statistics. In this section, we construct three scaling relations for galaxy clusters using the TNG300 and MTNG simulations: i) $L_X$ - $M_{\rm 500}$, ii)  $M_{\rm gas, \rm 500}$ - $M_{\rm 500}$, and iii) $k_B T$ - $M_{\rm 500}$. The X-ray luminosity is obtained by summing the contributions from all gas cells enclosed within $R_{500}$, while the cluster temperature corresponds to the mass-weighted temperature.  To construct these scaling relations, we employ the pipeline described in Section~\ref{sec:Mocks}. We consider only the halos with masses $M_{500} > 10^{12.5}\, \text{M}_{\odot}$ in TNG300 and $M_{500} > 10^{13.5}\, \text{M}_{\odot}$ in MTNG. To enhance the clarity of our results, we construct the $L_{\rm 500}$ - $M_{\rm 500}$ scaling relation by binning the clusters on a logarithmic mass scale, with each bin having a width of $0.1$ dex, considering a minimum of 6 galaxy clusters. For each mass bin, we calculate the average luminosity using the geometric mean (i.e.~mean in the log), as this is less sensitive to extreme values or outliers than the arithmetic mean, thereby providing a more reliable measure of the central trend \citep{Mitchell2004}:
\[
\bar{L}_{500} = \exp \left( \frac{1}{n} \sum^{n}_{i=1} {\ln}\, L_{500, i} \right).
\]
Here $L_{500,i}$ denotes the luminosity of the $i^{\rm\, th}$ cluster within a particular mass bin, and $n$ represents the total number of clusters in that bin. Additionally, the data dispersion is calculated using the geometric standard deviation, which is defined as:
\[
\sigma =\exp \sqrt{\frac{1}{n} \sum_{i=1}^{n} \left(\ln \frac{L_{500,i}}{\bar{L}_{500}}\right)^2 }.
\]
 In Figure~\ref{fig:M500_L}, we present the scaling relation obtained in this way for the TNG300 (blue) and MTNG (cyan) simulations. The data points represent mean values, while the error bars show their spread. For comparison, we also include observational data from different studies. Throughout this work, luminosities are displayed as $L_{500}\,E(z)^{-2}$, where $E(z)=H(z)/H_0$ is the dimensionless Hubble parameter. We adopt the $E(z)^{-2}$ normalisation because, in the soft X-ray band (0.5--2.0\,keV), the ICM emissivity is approximately independent of temperature over the cluster temperature range~\citep{Bohringer2012}, yielding a self-similar scaling $L_{X,\mathrm{soft}} \propto M\,E^2(z)$ rather than the $L_{X,\mathrm{bol}} \propto M^{4/3}\,E^{7/3}(z)$ expected for the bolometric luminosity.
 
 The scaling relations obtained from the TNG300 and MTNG simulations agree well with the observed datasets, showing that these simulations can accurately reproduce the mass-luminosity relationship observed specifically in the galaxy clusters region. The relatively small error bars, representing the geometrical standard deviation, are consistent with the spread seen in observational data. The overlap between the TNG300 and MTNG data points and the observational data indicates that the simulations consistently capture the physical processes governing the mass-luminosity relationship in galaxy clusters across a broad range of halo masses, from $10^{13.5}$ to $10^{15.5}$ $\text{M}_{\odot}$, and X-ray luminosities $L_{X}$, from $10^{42}$ to $10^{45}$ $\text{erg}\, s^{-1}$. This ensures that the observed trend is not limited to a specific subset of clusters, but rather reflects a general characteristic of galaxy clusters.

As illustrated in Figure~\ref{fig:M500_L}, observational data within the galaxy group regime ($M_{\rm 500} < 10^{14}\, \text{M}_{\odot}$) is relatively sparse, primarily because these systems possess shallow potential wells and faint X-ray emission that make them difficult to distinguish from the background. A more noticeable departure from self-similarity occurs near the $10^{14}\, \text{M}_{\odot}$ mass scale, where the scaling relation departs from the predictions of simple gravitational collapse models.  This transition is physically driven by the failure of the assumption that the ICM is heated exclusively by the gravitational potential of the system~\citep{Mantz2016b}. While gravity remains the dominant driver in high-mass clusters, the thermodynamics of lower-mass systems is significantly influenced by non-gravitational processes. Specifically, mechanisms such as AGN feedback and supernova-driven winds inject additional energy and entropy into the medium~\citep{Ansarifard2020, Marini2025, Molendi2025, Pratt2020, Toptun2025, Cucchetti2018}. 

In these lower-mass halos, the energy supplied by feedback can match or even exceed the gravitational binding energy, leading to a significant depletion of baryons as gas is expelled from the galaxy and redistributed toward the halo outskirts~\citep{Marini2025}. Furthermore, these feedback processes are responsible for the chemical enrichment of the ICM, as metals produced by stellar evolution are transported and mixed into the diffuse gas~\citep{Ansarifard2020}. Ultimately, the cumulative effect of these non-gravitational mechanisms breaks the scale-free nature of cluster formation, resulting in scaling relations that are significantly steeper than those predicted by purely gravitational models.

\begin{figure}
\centering
\includegraphics[width=\linewidth]{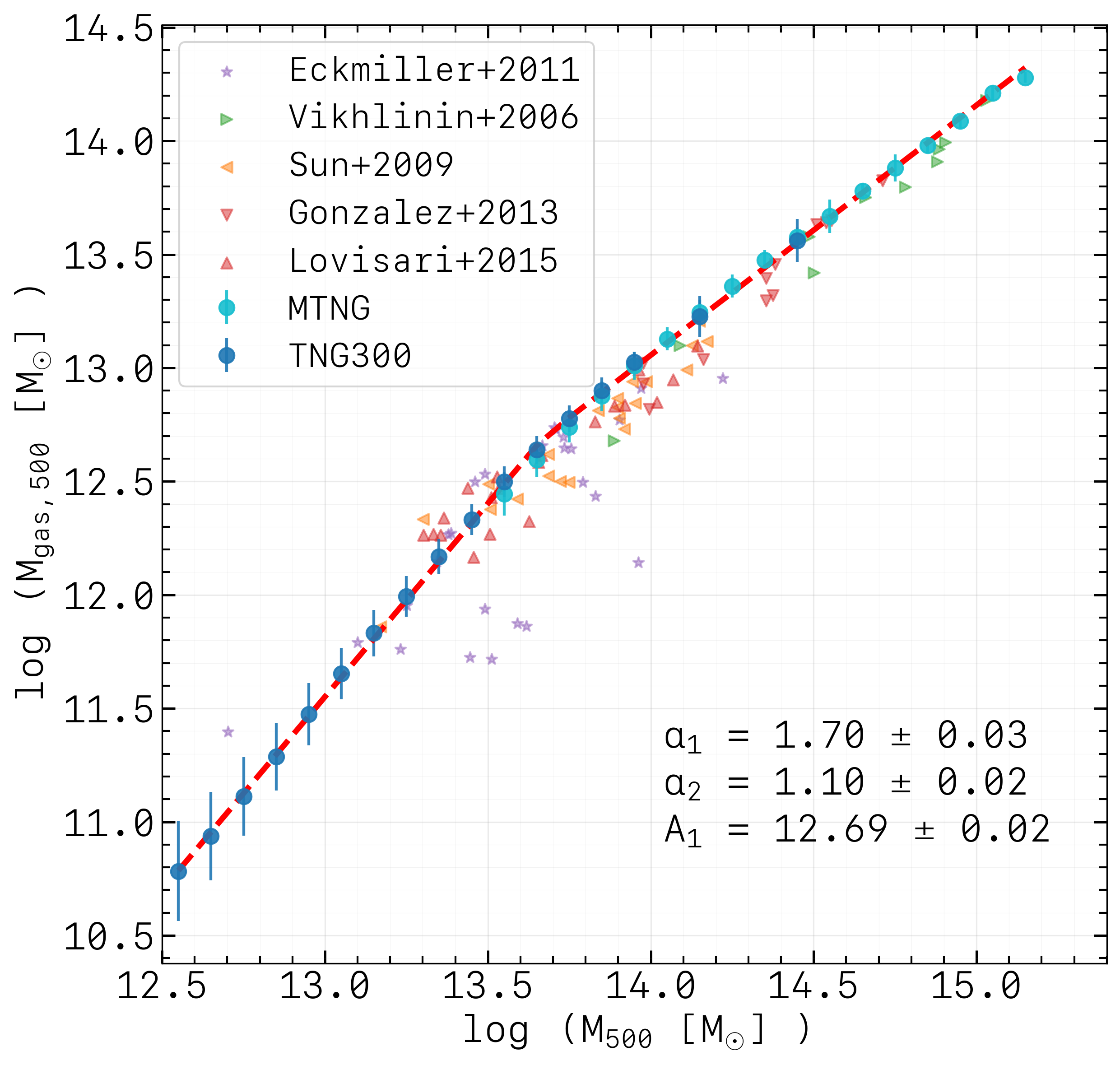}
\caption{Gas mass $M_{\mathrm{gas},500}$ as a function of total halo mass $M_{500}$ for the TNG300 (blue) and MTNG (cyan) simulations, compared with observational data. Both $M_{500}$ and $M_{\mathrm{gas},500}$ are computed by summing the masses of all particles and of gas particles, respectively, enclosed within $R_{500}$. Data points and error bars represent the geometric mean and geometric standard deviation within logarithmic mass bins. The solid lines show the best-fit broken power-law model. The simulated relation is in broad agreement with observational measurements across the full mass range, with the steepening at group scales reflecting the increasing efficiency of gas expulsion by AGN feedback in shallower potential wells.}
\label{fig:M500_MG500}
\end{figure}

The scaling relations derived from TNG300 and MTNG are consistent with a broken power law, with a transition mass $M_{\rm pivot}$ marking the scale at which the slope changes. Following~\citet{Brun2017}, we fit the $L_{X}$ versus $M_{\rm 500}$ relation using a broken power law of the form:

\[
L_{X} =
\begin{cases} 10^{A_1}\Big(\frac{{M}_{500}}{M_{\rm Pivot}}\Big)^{\alpha_1}, & \text{for}\;{M}_{500} < M_{\rm Pivot}, \\
10^{A_2}\Big(\frac{{M}_{500}}{M_{\rm Pivot}}\Big)^{\alpha_2}, & \text{for}\; {M}_{500} \geq M_{\rm Pivot}.
 \end{cases} 
\]

\begin{table*}[h]
\centering
      \caption{Best-fit parameters of the broken power-law model, applied to the three X-ray scaling relations derived from the combined TNG300 and MTNG samples. The fitted pivot mass is $M_{\rm pivot} = 10^{13.67}\,\mathrm{M}_\odot$ for the $L_X$--$M_{500}$ relation; this value is subsequently fixed for the $M_{\mathrm{gas},500}$--$M_{500}$ and $k_{\rm B}T$--$M_{500}$ fits. $\alpha_1$ and $\alpha_2$ denote the logarithmic slopes below and above the pivot mass, respectively, and $A_1$ is the normalisation at the pivot. The self-similar predictions for each relation are listed for comparison.}
         \label{Tab:Fitting}
\begin{tabular}{l c c c c} 
 \hline \hline
Scaling Relation &$\alpha_1$ &$\alpha_2$ &$A_1$ & Self-similarity expected $\alpha$ \\
 \hline
 $L_X$ - $M_{500}$ & 2.47 $\pm$ 0.37 &1.47$\pm$ 0.13 & 42.93$\pm$ 0.49 & 1.0 (for the X-rays soft band)\\
 $M_{gas, 500}$ - $M_{500}$ & 1.70 $\pm$ 0.03 & 1.10$\pm$ 0.02& 12.69$\pm$ 0.02& 1.0 \\
 $k_{B}T$ - $M_{500}$ & 0.63 $\pm$ 0.03& 0.56$\pm$ 0.02& 0.03$\pm$ 0.02 & 0.66\\
 \hline
\end{tabular}
\end{table*}

Since the fitted function must be continuous at $M_{\rm Pivot}$, we impose the additional constraint that $A_1 = A_2$. This allows us to fit the scaling relation with four free parameters: $A_1$, $\alpha_1$, $\alpha_2$, and $M_{\rm Pivot}$. Given that the MTNG and TNG300 simulations use the same physical model and yield similar X-ray luminosity values for clusters with equivalent mass, and to ensure robust statistics across the $10^{12.5} - 10^{15} $ mass range, we combined the samples from both simulations for the fitting process. The fitted parameters, along with their confidence intervals, are summarized in Table~\ref{Tab:Fitting}. Additionally, the slopes of this scaling relation and the fitted model are presented in Figure~\ref{fig:M500_L}.

This fitting approach provides a natural separation between 
low- and high-mass systems through the pivot mass $\log_{10}(M_{\rm pivot}/\mathrm{M}_\odot) = 13.67 \pm 0.27$, which marks the scale at which the slope of the scaling relation changes. For the subsequent $M_{500}$--$M_{\mathrm{gas},500}$ and $M_{500}$--$k_{\rm B}T$ relations, $M_{\rm pivot}$ is fixed to this value. For low-mass clusters ($M_{\rm 500} < 10^{13.67}\,\text{M}_{\odot}$), the logarithmic slope derived from our simulations is $\alpha_1 = 2.47 \pm 0.37$, reflecting the significant impact of non-gravitational processes such as AGN feedback and radiative cooling, which steepen the relation in shallower potential wells. In contrast, for high-mass clusters ($M_{\rm 500} \geq 10^{13.67}\, \text{M}_{\odot}$), the logarithmic slope is $\alpha_2=1.47 \pm 0.13$, a value that is closer to the slope of $1$ predicted by the purely gravitational self-similar model for the soft X-ray emission. This indicates that at the highest mass scales, clusters behave as systems where gravitational collapse is the primary driver of the thermodynamic state of the gas. The inferred value of \(M_{\rm pivot}\) is close to the mass scale commonly adopted to distinguish the cluster and group regimes, typically \(10^{14}\,M_{\odot}\). For the subsequent scaling relations, \(M_{500}\)–\(M_{{\rm gas},500}\) and \(M_{500}\)–\(k_{\rm B}T\), we fix \(M_{\rm pivot}\) to the value derived from this relation. 

Observational determinations of the $L_{\mathrm{X}}$--$M_{500}$ relation in the soft X-ray band generally report slopes in the range $\sim 1.3 - 1.6$ for massive clusters. For example, \citet{Eckmiller2011}, \citet{Vikhlinin2009}, \citet{Mantz2016b}, and \citet{Maughan2007} derived slopes of $1.44 \pm 0.10$, $1.61 \pm 0.14$, $1.34 \pm 0.05$, and $1.45 \pm 0.07$, respectively, for cluster samples spanning masses $\gtrsim 10^{14}\, \text{M}_{\odot}$. A steeper value was reported by \citet{Zhang2008}, who found a slope of $2.33 \pm 0.70$ (for the bolometric luminosity) using 37 XMM-Newton clusters with masses between $2 \times 10^{14}$ and $10^{15}\, \text{M}_{\odot}$. In the lower-mass regime, \citet{Lovisari2015} measured a slope of $1.66 \pm 0.22$ for galaxy groups, ~\citet{Eckmiller2011} derived a value of $1.34 \pm 0.18$, while \citet{Reiprich2002} obtained $1.61 \pm 0.09$ from an extended sample covering masses from $3 \times 10^{13}$ to $3 \times 10^{15}\, \text{M}_{\odot}$.

A particularly valuable comparison in the group regime is provided by the recent eFEDS analysis of \citet{Popesso2025}, whose stacked measurements are included in Figure~\ref{fig:M500_L}. Their sample is drawn from the optically selected GAMA galaxy group catalogue, and the average X-ray luminosities are obtained through a stacking analysis of the eROSITA eFEDS data that reaches from poor clusters down to Milky Way-like halos ($M_{500} = 10^{12.5}\,\text{M}_{\odot}$). Because the sample is optically rather than X-ray selected, it is far less affected by the Malmquist-type selection biases that hamper flux-limited group catalogues at fixed halo mass, and is therefore closer in construction to our volume-limited simulation sample. As shown in Figure~\ref{fig:M500_L}, our simulated $L_{\mathrm{X}}$--$M_{500}$ relation is in good agreement with the \citet{Popesso2025} measurements throughout the group regime, where these data populate the mass range in which the other observational samples are sparse and follow the same trend as the TNG300 and MTNG points down to the lowest masses probed. The agreement is tightest at the Milky Way-like scale, while at the more massive group and poor-cluster scale our relation lies marginally above their stacked points, although the offset stays within the combined uncertainties. This consistency with an optically selected, weakly biased sample that extends deep into the group regime reinforces our interpretation that the departure from self-similarity at group scales is a genuine physical signature of non-gravitational feedback rather than an artifact of the X-ray selection or analysis.

In Figure~\ref{fig:M500_MG500}, we present the scaling relation between the total halo mass $M_{500}$ and the gas mass $M_{\mathrm{gas},500}$ for the TNG300 and MTNG simulations. Here, $M_{500}$ is defined as the sum of the masses of all particles enclosed within $R_{500}$, while $M_{\mathrm{gas},500}$ is the corresponding sum over gas particles only. The spread, indicated by the bars, is narrower than that observed in the luminosity scaling relation. Our simulated results demonstrate broad consistency with observational data across a wide mass range, though the dispersion reported by \citet{Eckmiller2011} is significantly larger than both our predictions and other literature samples. This heightened scatter in the \citet{Eckmiller2011} group sample arises from observational challenges in the low-temperature regime, such as limited radial coverage or inaccuracies in background modelling and gas mass calculations for faint structures. 
To characterize the departure from the purely gravitational self-similar prediction $(\alpha=1)$, and following the fitting done for the $L_{\rm 500} - M_{\rm 500}$ scaling relation, we modelled the relation using a broken power law. We obtained logarithmic slopes of $\alpha_1= 1.70 \pm 0.03$ for the low-mass regime and $\alpha_2 = 1.10 \pm 0.02$ for high-mass clusters using the same mass pivot of $M_{500} = 10^{13.67}\, \text{M}_{\odot}$. As before, the steepening observed at lower masses indicates that non-gravitational processes, primarily AGN kinetic feedback, are more efficient at expelling gas from the shallower potential wells of galaxy groups compared to massive clusters~\citep{Marini2025, Molendi2025}. In contrast, the slope for high-mass clusters ($\alpha_2 = 1.1 \pm 0.02$) approaches the self-similar expectation, as these systems behave as closed boxes where gravity dominates the energy budget~\citep{Mantz2016b, Bohringer2012}. This simulated high-mass slope is in good agreement with the observational value of $0.906 \pm 0.08$ reported by \citet{Zhang2008}.

\begin{figure}
\centering
\includegraphics[width=\linewidth]{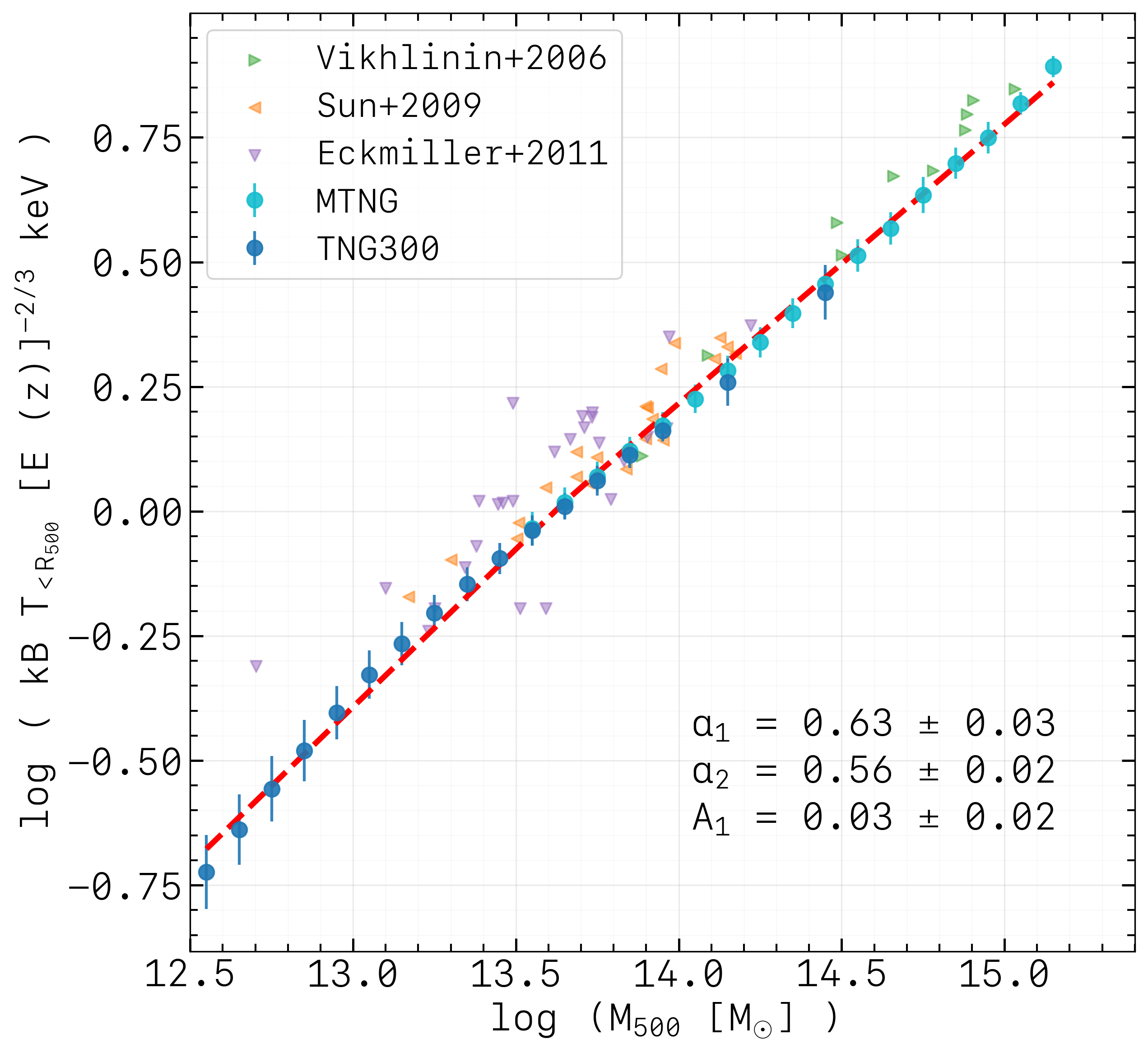}
\caption{Mass-weighted temperature $k_{\rm B}T$ as a function of halo mass $M_{500}$ for the TNG300 (blue) and MTNG (cyan) simulations, compared with observational data. The temperature is defined as the mass-weighted mean over all hot, non-star-forming gas cells within $R_{500}$. Data points and error bars represent the geometric mean and geometric standard deviation within logarithmic mass bins. The solid lines show the best-fit broken power-law model. The simulated temperatures are in overall agreement with observations, though a systematic offset of approximately $0.1\,\mathrm{dex}$ toward lower values is present, likely reflecting differences between the mass-weighted temperature used in the simulations and the spectroscopic temperature recovered by X-ray detectors.}
\label{M500_T}
\end{figure}

Figure~\ref{M500_T} presents the scaling relation between the mass-weighted temperature and the cluster mass \(M_{\rm 500}\). When compared to observational data, there is a slight underprediction of the temperature of approximately \(0.1 \text{ dex}\). We also fitted this scaling relation using a broken power law model, with the fitting parameters presented in the last row of Table~\ref{Tab:Fitting}. The logarithmic slopes found for this scaling relation are $\alpha_1= 0.63 \pm 0.03$ for low-mass clusters and $\alpha_2= 0.56\pm 0.02$ for high-mass clusters. These values are consistent with those derived observationally: \cite{Ettori2004} and \citet{Hicks2008} obtained a value of 0.58, while~\citet{Arnaud2005} derived a value of 0.56. 

A possible contribution to the temperature offset is the difference in how temperatures are defined in simulations and in X-ray observations. In simulations we adopt the mass-weighted temperature, which traces the true thermal energy content of the intracluster gas, whereas observational temperatures are derived from spectral fitting and thus correspond to a projected, spectroscopic quantity. Although more refined estimators, such as the spectroscopic-like temperature introduced by \citet{Mazzotta2004} and then generalized by \citet{Vikhlinin2006}, were specifically designed to approximate the temperature recovered by X-ray detectors, they are intrinsically biased toward the densest and coolest gas along the line of sight. It means that temperatures derived through this method are systematically lower than both the mass-weighted and emission-weighted definitions, especially in clusters with multiphase structure or significant thermal gradients. Consequently, adopting the spectroscopic-like temperature would further decrease the simulated temperatures, amplifying the discrepancy with observations.

Unlike the $M_{\rm 500}$--$L_X$ and $M_{\rm 500}$--$M_{\rm gas, \rm 500}$ relations, which exhibit a distinct break between the galaxy group and cluster regimes, the mass--temperature ($M$--$T$) scaling relation remains remarkably consistent across both mass regimes, and the slopes are very close to each other. This finding aligns with recent observational analyses by \citet{Toptun2025}, who utilized stacked eROSITA spectra to measure the halo $M$-$T$ relation over two decades in mass. Their results indicate that a single power law (with a slope of $T \propto M^{0.6}$) provides an excellent fit for both groups and clusters, with no statistically significant break in the slope across the sampled range \citep{Toptun2025}. 

These results are in agreement with the self-similar prediction of $T \propto M^{0.66}$. This consistency suggests that the global gas temperature is primarily determined by the depth of the gravitational potential well \citep{Sarazin1986, Toptun2025}. Consequently, while non-gravitational feedback processes are effective at redistributing or ejecting gas, they have a comparatively minor impact on the integrated temperature profile \citep{Toptun2025}.

In summary, although this is not the first work to derive cluster scaling relations from simulations, it is noteworthy how well our results reproduce the observed trends. Using the TNG300 and MTNG simulations, we recover the observed scaling relations for galaxy clusters with high fidelity. In particular, the simulations closely match the \(L_{\rm X} - M_{\rm 500}\) and \(M_{\rm gas,\rm 500} - M_{\rm 500}\) relations. The \(k_B T - M_{\rm 500}\) relation shows a small offset, with simulated temperatures slightly lower than those measured observationally. The strong dependence of X-ray luminosity on the square of the gas density and the square root of the gas temperature makes the emission especially sensitive to the physical processes implemented in the simulations. Variations in gas properties, such as those induced by mergers or strong cooling episodes, naturally introduce scatter in the predicted luminosities, which is reflected in the larger dispersion observed in Figure~\ref{fig:M500_L} compared to the \(M_{\rm gas,\rm 500} - M_{\rm 500}\) relation.  The overall agreement between simulated and observed data indicates that the simulations broadly capture the key physical mechanisms that regulate X-ray emission, including gas density structure, thermal state, and the dynamical evolution of clusters. Moreover, the TNG300 and MTNG simulations reproduce the expected slopes for massive clusters (\(M_{\rm 500} > 10^{13.67} \, \text{M}_{\odot}\)) particularly well for both the luminosity and gas-mass relations, and they remain reasonably close for the temperature relation. The inclusion of MillenniumTNG is key to establishing these trends at the high-mass end ($M_{500} \geq 10^{14.5}\,\mathrm{M}_\odot$), a regime statistically inaccessible in TNG300 alone, and the metallicity-based prescription developed here enables, for the first time, consistent X-ray analysis of a simulation that does not track individual chemical abundances.

\section{Mass and luminosity derivation from synthetic X-ray maps}
\label{Sec:Mass_baias}
\begin{figure}
\centering
\includegraphics[width=\linewidth]{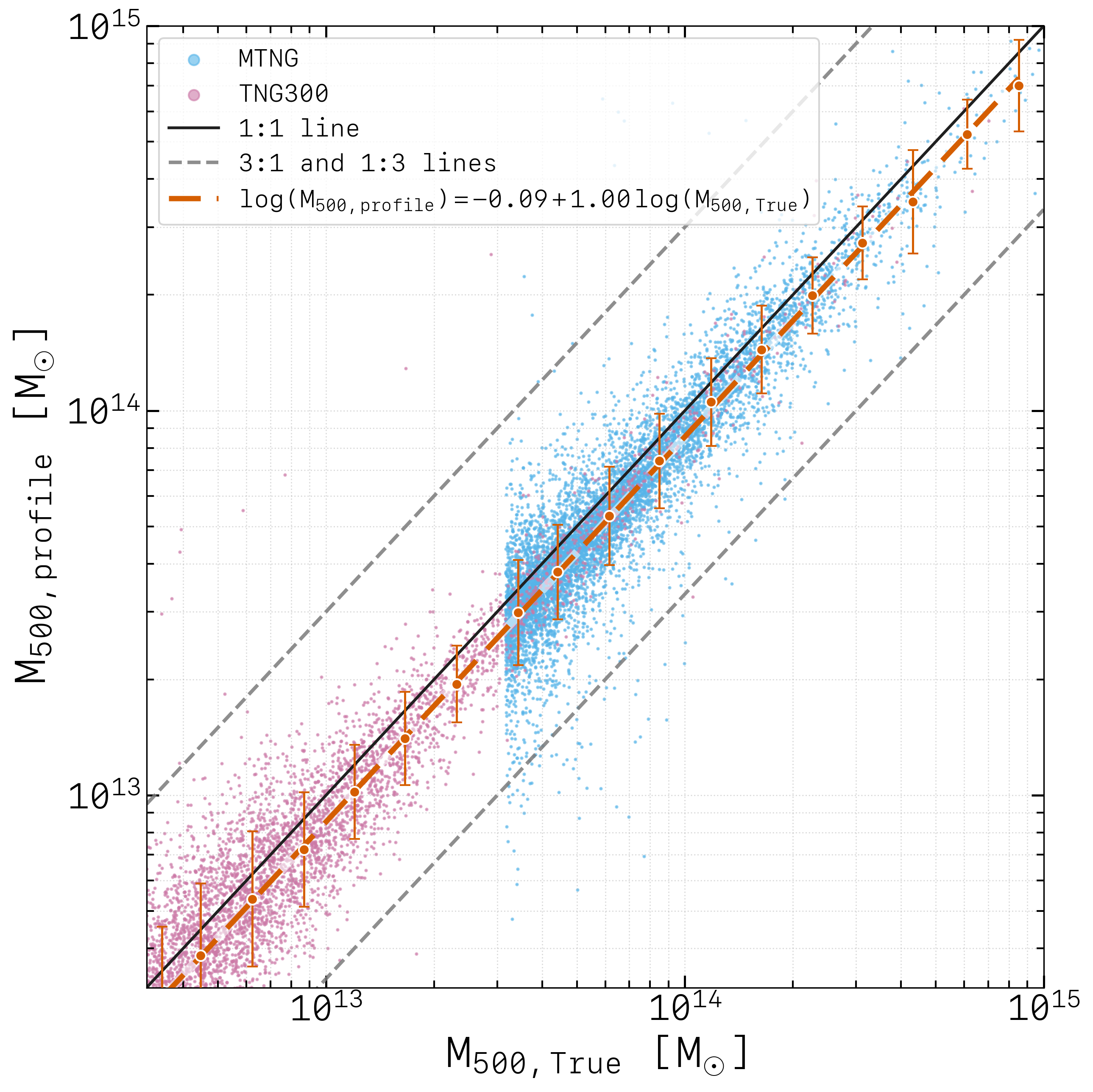}
\caption{Comparison between the true halo mass $M_{500,\mathrm{true}}$, defined as the sum of all particle masses within $R_{500}$, and the hydrostatic equilibrium mass $M_{500,\mathrm{profile}}$ inferred from 3D density and temperature profiles extracted directly from the simulation, excluding the central region ($R < 0.15\,R_{500}$). The hydrostatic mass is systematically underestimated by approximately $15\%$ across the full mass range studied, demonstrating that the hydrostatic equilibrium assumption alone introduces a mass-independent systematic bias even when the gas profiles are known exactly.}
\label{Fig:profile_true}
\end{figure}

Up to the last section, the physical properties and scaling relations of galaxy clusters have been derived directly from simulation outputs. By utilizing the full three-dimensional thermodynamic and dynamical phase-space information, the true cluster mass ($M_{\rm 500, \rm true}$) was determined by summing the individual masses of all particles or volume elements enclosed within a specific radius, such as $R_{\rm 500}$. Similarly, the X-ray luminosity ($L_X$) was computed by integrating the emissivity of all hot, non-star-forming gas elements based on their local density, temperature, and metallicity using collisional ionization equilibrium models like APEC.

In contrast, the reconstruction of galaxy cluster properties from observations must rely on indirect methods and a series of important assumptions \citep{Biffi2012}. A primary requirement is the assumption that the ICM is in a state of hydrostatic equilibrium (HSE) within a spherically symmetric gravitational potential \citep{Sarazin1986, Vikhlinin2006, Kravtsov2006}. Within this framework, the inward gravitational force exerted by the total mass (dominated by dark matter) is assumed to be perfectly balanced by the outward thermal pressure gradient of the gas~\citep{Angelinelli2020, Braspenning2025}. Furthermore, spherical symmetry implies that the cluster’s mass distribution and gas properties depend exclusively on the radial distance from the center and are uniform in all directions~\citep{Andrade-Santos2017, Kolokotronis2001, Lau2009}. While these assumptions facilitate mass estimation from projected X-ray profiles, they often neglect non-thermal pressure support, such as turbulence and bulk motions, which can lead to a systematic errors in the mass estimation~\citep{Angelinelli2020, Lau2009, McNamara2007}. In this section, we investigate how the assumption of hydrostatic equilibrium and spherical symmetry affect mass estimates and scaling relations by applying observationally motivated procedures to the simulated data.

To isolate the systematic effects of fundamental assumptions on cluster mass reconstruction, we compare the true halo mass ($M_{\rm 500,\rm true}$) with two observationally motivated mass estimates. The true mass is obtained by summing the individual masses of all particles or volume elements enclosed within the three-dimensional radius $R_{\rm 500}$. To derive the first estimate, denoted as $M_{\rm 500,\rm profile}$, we construct intrinsic 3D radial profiles by partitioning each cluster into 25 linearly spaced spherical shells. We exclude the central regions of galaxy clusters ($R < 0.15\,R_{500}$), a procedure commonly adopted to mitigate the influence of non-gravitational physics, such as radiative cooling and AGN feedback, which govern the thermodynamic state of cluster cores~\citep{Eckmiller2011, Toptun2025, Allen2011}. Since X-ray emissivity scales with the square of the gas density, the presence of dense, centrally peaked cool cores can substantially boost the integrated luminosity, introducing scatter and selection biases in cluster samples~\citep{Andrade-Santos2017, Braspenning2025, Lovisari2015, Mantz2010, Allen2011}. In the present work, this exclusion additionally mitigates the absence of an interstellar medium contribution and the omission of AGN emissivity in our pipeline.

Within each shell, we compute the mass-weighted temperature ($T_{\rm mw}$) and mean gas density directly from the simulation outputs. These data are subsequently fitted with the parametric functional forms proposed by \citet{Vikhlinin2006} (Equations~\ref{Eq:temp_profile} and \ref{Eq:density}), which provide the smooth derivatives necessary to solve the hydrostatic equilibrium equation (Eq.~\ref{Hydrostatic_equilibrium_equation}). The second estimate, $M_{\rm 500,\rm spec}$, is derived by applying the same analytical framework to density and temperature profiles extracted from synthetic X-ray emission maps, assuming a single temperature and density model for each bin. This mock pipeline accounts for the quadratic density dependence of the X-ray emission and utilizes  APEC spectral fitting to mimic the temperatures typically recovered by X-ray detectors. By contrasting $M_{\rm 500, \rm profile}$ and $M_{\rm 500,\rm spec}$ against the true simulation mass, we can precisely quantify the hydrostatic mass bias and determine how much of the error is due to true physical deviations from equilibrium versus biases inherent in the X-ray measurement process. We emphasise that our pipeline generates ideal emission maps rather than fully realistic mock observations: no instrument response matrix, effective area curve, redistribution function, or photon noise is applied. The spectral fitting in each radial bin consists of matching a single-temperature, single-metallicity APEC model to the volume-integrated theoretical spectrum of that bin over the 0.5--2.0\,keV band, without any instrumental degradation. The spectra are tabulated on a grid of 2\,000 linearly spaced energy bins across this band, corresponding to an energy resolution of $\Delta E = 0.75$\,eV. This approach isolates the biases arising from the spectral modelling assumptions (single temperature and metallicity per radial bin) and from the hydrostatic equilibrium assumption, without convolving them with photon statistics or instrumental calibration uncertainties. A more realistic treatment including telescope response convolution and photon counting noise would introduce additional scatter but is not expected to qualitatively alter the systematic offsets reported here

\begin{figure}
\centering
\includegraphics[width=\linewidth]{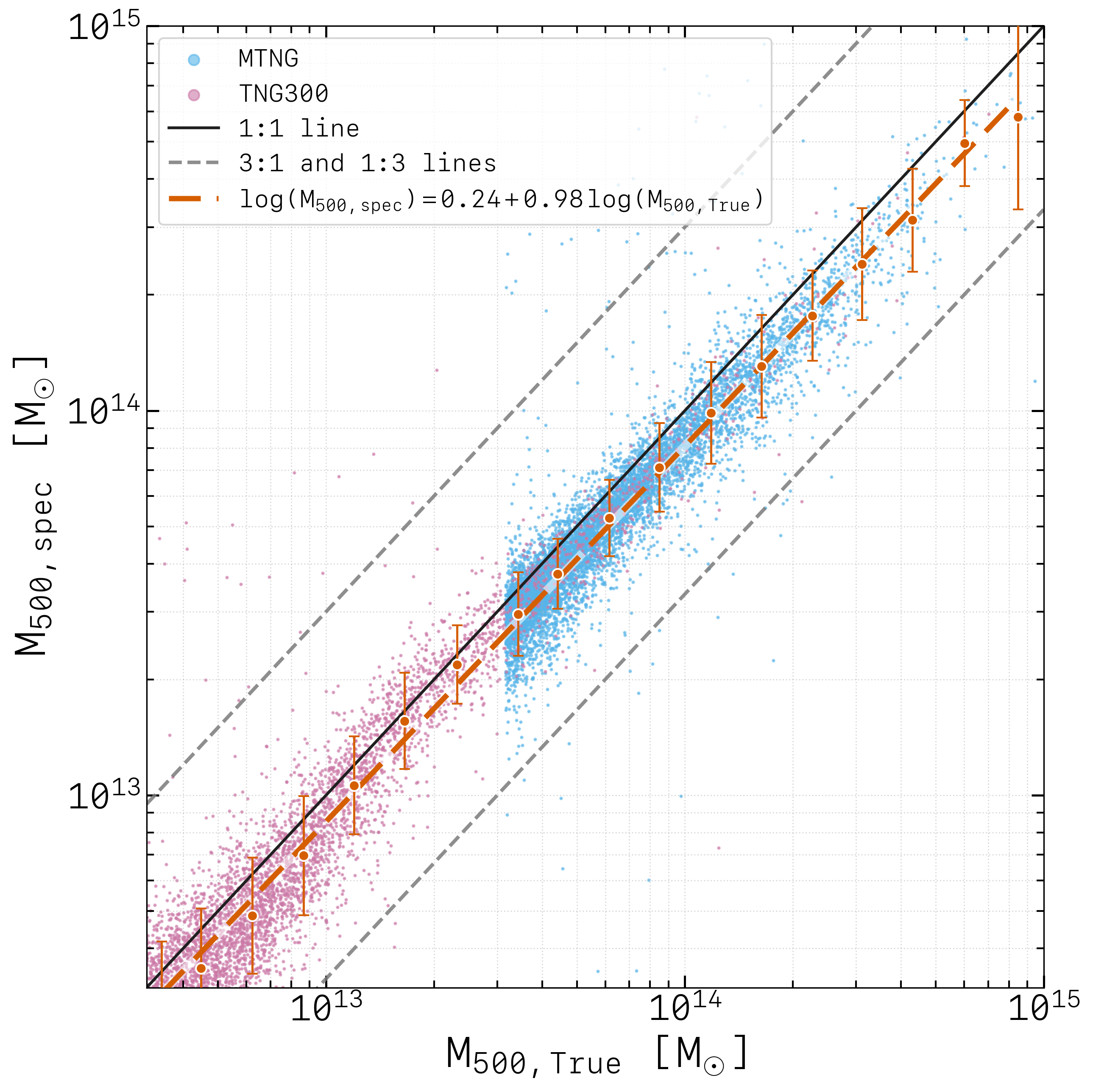}
\caption{Comparison between the true halo mass $M_{500,\mathrm{true}}$ and the spectroscopic mass $M_{500,\mathrm{spec}}$ derived from the mock X-ray analysis pipeline, excluding the central cluster region ($R < 0.15\,R_{500}$). The one-to-one relation is shown by the dashed line. Unlike the profile-based hydrostatic mass bias, the spectroscopic mass offset is mass-dependent: low-mass clusters are underestimated by approximately $15\%$, while the discrepancy increases to approximately $21\%$ for the most massive systems, reflecting the growing impact of spectroscopic temperature biases in multi-temperature gas at higher masses.}
\label{Fig:spec_true}
\end{figure}

Figure~\ref{Fig:profile_true} presents a comparison between the true halo mass ($M_{500, \rm true}$)  and the hydrostatic mass inferred from simulation-extracted density and temperature profiles. We identify a systematic offset where the hydrostatic equilibrium (HSE) framework underestimates $M_{500, \rm profile}$ by approximately 15\% relative to the simulation ground truth in all of the studied mass range. This result is particularly noteworthy as it demonstrates that even with perfectly known gas profiles (free from instrumental noise or background contamination), the fundamental assumption of hydrostatic balance alone leads to a systematically biased mass estimate.
This $\sim$ 15\% bias aligns with the well-documented hydrostatic mass bias observed in both real and simulated clusters, where HSE-derived masses typically fall 10\%–30\% below the true gravitating mass~\citep{Lau2009, Ansarifard2020, Braspenning2025}. The physical origin of this discrepancy lies in the presence of non-thermal pressure support generated by subsonic gas motions, including turbulence, rotation, and large-scale bulk flows. Because the standard HSE equation assumes that the inward pull of gravity is balanced exclusively by the outward thermal pressure gradient, it neglects the additional support provided by these kinetic components~\citep{Lau2009, Allen2011}.
Furthermore, simulations indicate that this bias is not spatially uniform; it tends to increase with cluster-centric radius, as the gas in the outskirts is often less relaxed and has had less time to fully thermalize following accretion from the cosmic web~\citep{Lau2009, Pratt2019}. These residual motions are a result of the hierarchical assembly process, where mergers and continuous accretion drive persistent vorticity and shocks in the ICM~\citep{Barnes2021}. Consequently, our findings reinforce the conclusion that the HSE assumption introduces a systematic error in mass determination that exists independently of, and is often compounded by, further observational challenges such as projection effects or spectroscopic weighting biases. Importantly, this $15\%$ intrinsic bias is mass-independent, confirming that the steepening of the scaling relations in the group regime cannot be attributed to the hydrostatic equilibrium assumption alone, but is instead driven by the physical impact of baryonic feedback.

\begin{figure}
\centering
\includegraphics[width=\linewidth]{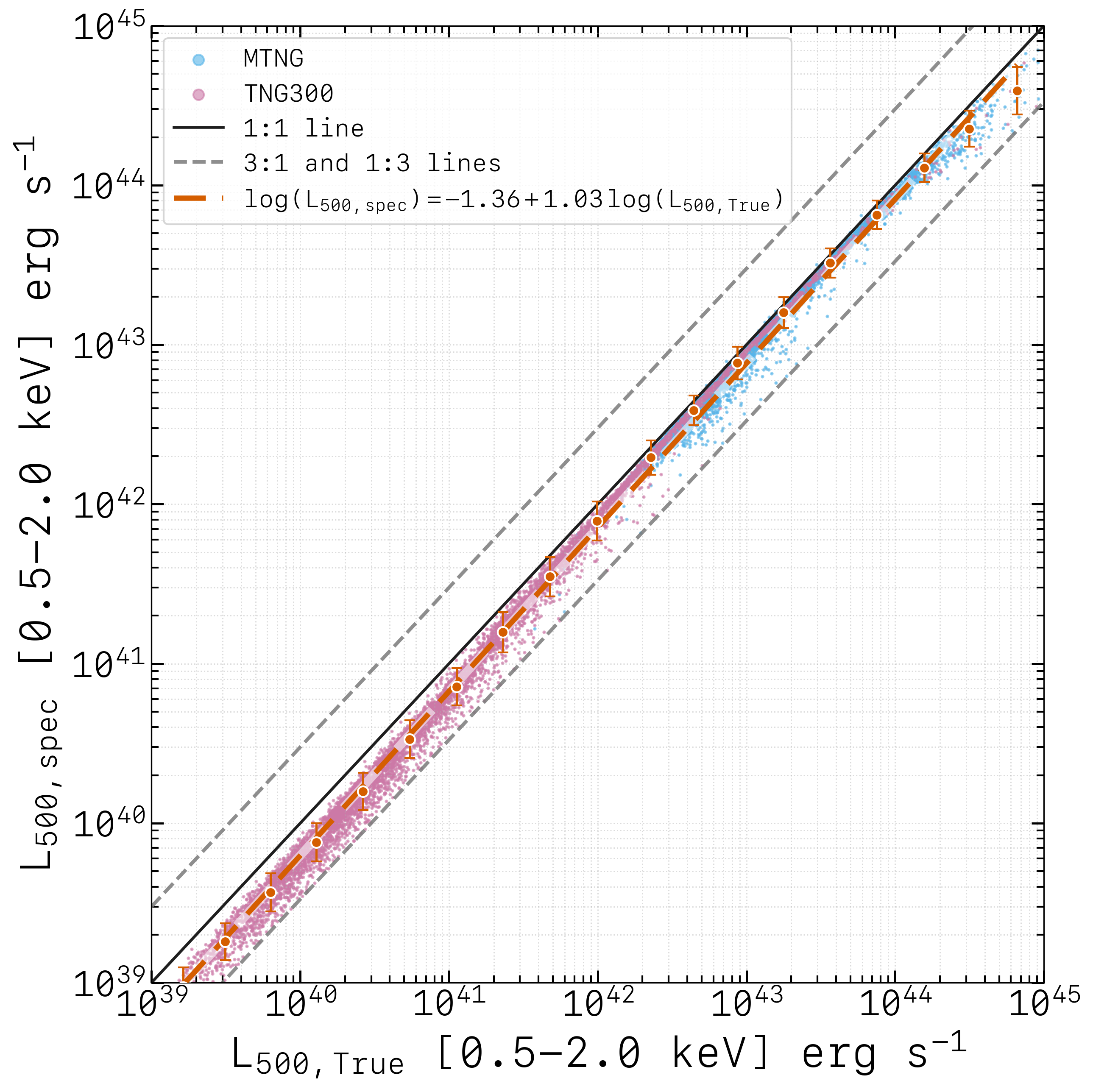}
\caption{Comparison between the true X-ray luminosity $L_{500,\mathrm{true}}$ and the spectroscopic luminosity $L_{500,\mathrm{spec}}$ recovered from spectral fitting of the synthetic X-ray maps. To ensure a like-for-like comparison, both quantities are computed from the X-ray emission in the $0.15$--$1\,R_{500}$ annulus, i.e.\ excluding the central cluster region ($R < 0.15\,R_{500}$). The luminosity offset is mass-dependent: high-mass clusters are underestimated by approximately $18\%$ on average, while the discrepancy increases to approximately $33\%$ for low-mass systems. This behaviour reflects the limitations of single-temperature, single-metallicity spectral modeling of the lower-density, multiphase gas outside the core, which become more severe in low-mass systems.}
\label{Fig:Xrays_true_spec}
\end{figure}

In Figure~\ref{Fig:spec_true}, we present the relationship between the true mass and the reconstructed spectroscopic mass ($M_{\rm 500,\rm spec}$). The results indicate that the offset between the spectroscopic and true mass is not uniform across the mass range, but exhibits a mass dependence. At the low-mass end, the spectroscopic mass is underestimated by approximately $15\%$ relative to the true mass, while at the high-mass end this discrepancy increases to approximately $21\%$. This mass-dependent behaviour contrasts with the approximately uniform $15\%$ offset found when using simulation-extracted profiles directly (Figure~\ref{Fig:profile_true}), and indicates that spectroscopic temperature biases contribute an additional, mass-dependent component to the total mass underestimation. Because X-ray emissivity scales with the square of the gas density, the emission measure is dominated by the coolest and densest gas phases within the multi-temperature ICM, causing the single-temperature fit to underestimate the mass-weighted temperature of the plasma. This causes the inferred spectroscopic temperature to be lower than the true mass-weighted temperature, directly propagating into a lower mass estimate~\citep{Rasia2006, Barnes2021, Zhang2024sca}. Beyond the systematic offset, the standard observational procedure introduces considerable scatter in the mass measurements. This scatter is intrinsically linked to the dynamical state of the cluster; for example, merging clusters often possess unthermalized bulk motions and temperature inhomogeneities that deviate from the mean scaling relations~\citep{Mantz2016b, Ettori2020, Farahi2019}. Recent studies suggest that identifying and accounting for the dynamical state, using indicators such as the centroid shift, symmetry, or concentration index, could reduce this scatter and refine the accuracy of mass proxies in future \mbox{X-ray} surveys~\citep{Yuan2020, Lovisari2020, Mantz2016a}. To further mitigate these systematic errors, recent studies have explored empirical corrections to the mass scale based on the shape of the reconstructed profiles. Specifically, \citet{Ansarifard2020} demonstrated that a robust correction to the hydrostatic mass bias can be inferred by combining X-ray measurements of gas inhomogeneity with the gradient of the gas density or pressure profiles. 

\begin{figure}
\centering
\includegraphics[width=\linewidth]{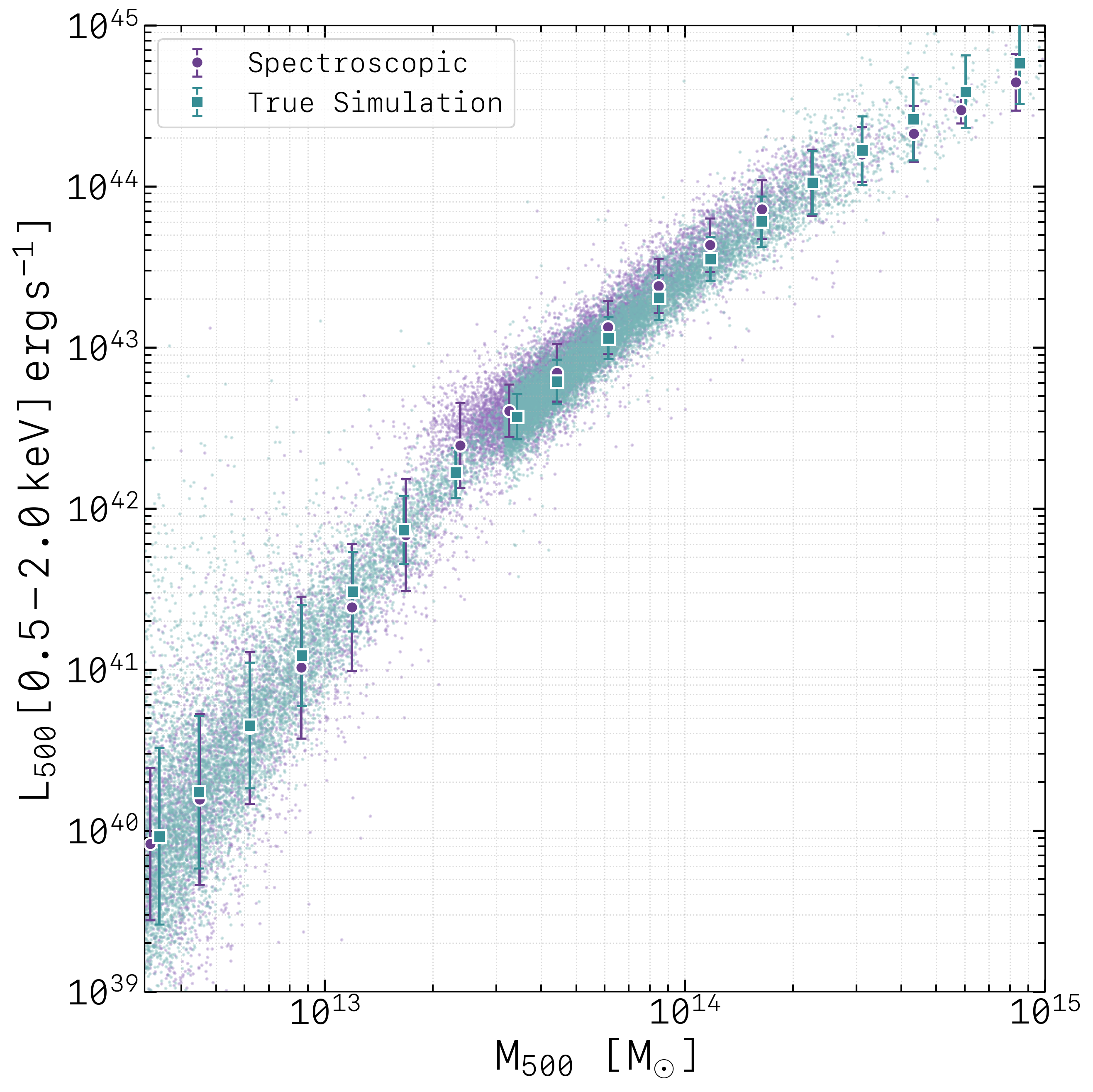}
\caption{Comparison of the $M_{500}$--$L_{500}$ scaling relation derived from two different definitions of the underlying quantities, both excluding the central cluster region ($R < 0.15\,R_{500}$). Green points show the simulation-based relation 
$(M_{500,\mathrm{true}},\,L_{500,\mathrm{true}})$, in which both 
quantities are computed directly from the particle data. Purple points 
show the spectroscopic relation $(M_{500,\mathrm{spec}},\,L_{500,\mathrm{spec}})$, derived from the mock X-ray observational pipeline. Despite the systematic and 
mass-dependent biases in the individually recovered quantities, the two scaling relations follow a consistent trend across the full mass range, demonstrating that the spectroscopic pipeline shifts mass and luminosity simultaneously, absorbing a fraction of the bias along the intrinsic relation rather than perpendicular to it.}
\label{Fig:sim-spec}
\end{figure}

In Figure~\ref{Fig:Xrays_true_spec}, we compare the X-ray luminosity $L_{500}$ measured directly from the simulation, obtained by summing the emission from hot gas particles, with the corresponding $L_{500,\rm spec}$ derived from spectral fitting of the synthetic X-ray maps. To ensure a like-for-like comparison, both luminosities are computed from the X-ray emission in the $0.15$--$1\,R_{500}$ annulus, excluding the core. The deviation of the spectroscopic value with respect to the simulation-based value is mass dependent. For high-mass clusters, $L_{500,\mathrm{spec}}$ is underestimated by approximately $18\%$ on average, whereas for low-mass systems this discrepancy increases to approximately $33\%$. These offsets exceed those found for the mass reconstruction because, once the dense central regions that dominate the emissivity budget are excluded, the recovered luminosity is governed by the lower-density, multiphase gas of the cluster outskirts, where the single-temperature, single-metallicity fit is least accurate. The mass dependence is consistent with the limitations of a single-temperature, single-metallicity model applied to each radial bin. In high-mass clusters ($k_{\mathrm{B}}T \gtrsim 2\,\mathrm{keV}$), the spectrum is dominated by bremsstrahlung emission, whereas line emission contributes more substantially in low-mass systems ($k_{\mathrm{B}}T \lesssim 2\,\mathrm{keV}$)~\citep{Mulchaey2000, Voit2005, Eckmiller2011}. Combined with the reduced photon counts and the increased influence of non-gravitational processes in the low-mass regime, these factors make the single-temperature and single-metallicity assumption more important, resulting in a larger bias in the recovered X-ray luminosity for low-mass clusters. The explicit quantification of the spectroscopic luminosity bias as a function of cluster mass, and its distinction from the mass bias, is a novel aspect of the present analysis not previously characterised within the TNG model.

Finally, Figure~\ref{Fig:sim-spec} compares the halo mass and X-rays luminosity ($M_{500}$-$L_{500}$) scaling relations derived using two definitions of the underlying quantities: the true simulation-based relation, $(M_{500,\mathrm{true}},\,L_{500,\mathrm{true}})$, shown in green, and the spectroscopic relation, $(M_{500,\mathrm{spec}},\,L_{500,\mathrm{spec}})$, shown in purple. The two relations follow a consistent trend across the full mass range considered, indicating that, although the spectroscopic pipeline introduces biases in the individually recovered quantities, the resulting scaling relation remains close to that obtained directly from the simulation. 

This behaviour can be understood in terms of the covariance between the biases in the inferred mass and luminosity. Because both observables are shifted simultaneously by the spectroscopic pipeline, a fraction of the bias is absorbed along the direction of the scaling relation rather than perpendicular to it, leaving the normalization and slope of the $M_{500}$-$L_{500}$ relation only weakly affected. This highlights that agreement at the level of the scaling relation does not imply unbiased recovery of the underlying physical quantities, but rather reflects the correlated manner in which the spectroscopic procedure modifies both variables~\citep{Barnes2021}.

From a physical perspective, the correlated shifts in the recovered mass and luminosity arise because both quantities are sensitive to the thermodynamic structure of the ICM, albeit through different dependencies. The X-ray luminosity scales with the square of the gas density, making it particularly sensitive to the density distribution and to the thermal and chemical state of the gas. The mass inferred under the assumption of hydrostatic equilibrium, by contrast, depends on the logarithmic gradients of both the density and temperature profiles (see Eq.~\ref{Hydrostatic_equilibrium_equation}), and is therefore sensitive to the accuracy with which the spectral fitting recovers the radial thermodynamic structure. Any simplification in the spectral modeling, such as the single-temperature and single-metallicity approximation applied per radial bin, propagates simultaneously into both $L_{500,\mathrm{spec}}$ and $M_{500,\mathrm{spec}}$, introducing correlated biases in the two observables. In low-mass systems, where line emission contributes substantially to the spectrum and the gas is more susceptible to cooling and feedback processes, the ICM departs more significantly from a single-phase description. The cooling function $\Lambda(T, Z)$ is strongly dependent on metallicity in this regime, and the spectral degeneracy between temperature, metallicity, and normalisation is more pronounced. These effects combine to produce those observed biases in the recovered luminosity and mass. In high-mass clusters, the spectrum is dominated by bremsstrahlung emission, for which $\Lambda(T) \propto T^{1/2}$~\citep{Sarazin1986, Voit2005}, reducing the sensitivity of the fit to line-complex modeling and yielding a more stable luminosity recovery. Nevertheless, the mass estimate remains sensitive to the recovered temperature gradient and to the validity of the hydrostatic equilibrium assumption, which may be violated in the presence of residual non-thermal pressure support.

It is worth emphasizing that the recovery of both the slope and normalization of the scaling relation is achieved because we exclude the central regions of galaxy clusters. However, in Appendix~\ref{Sec:Mass_baias_todo}, we show that when the luminosity and mass are derived for the entire halo, the slope remains consistent at the high-mass end, while the normalization exhibits a systematic shift.

\section{Redshift evolution}
\label{Sec:Redshift}
\begin{figure}
\centering
\includegraphics[width=\linewidth]{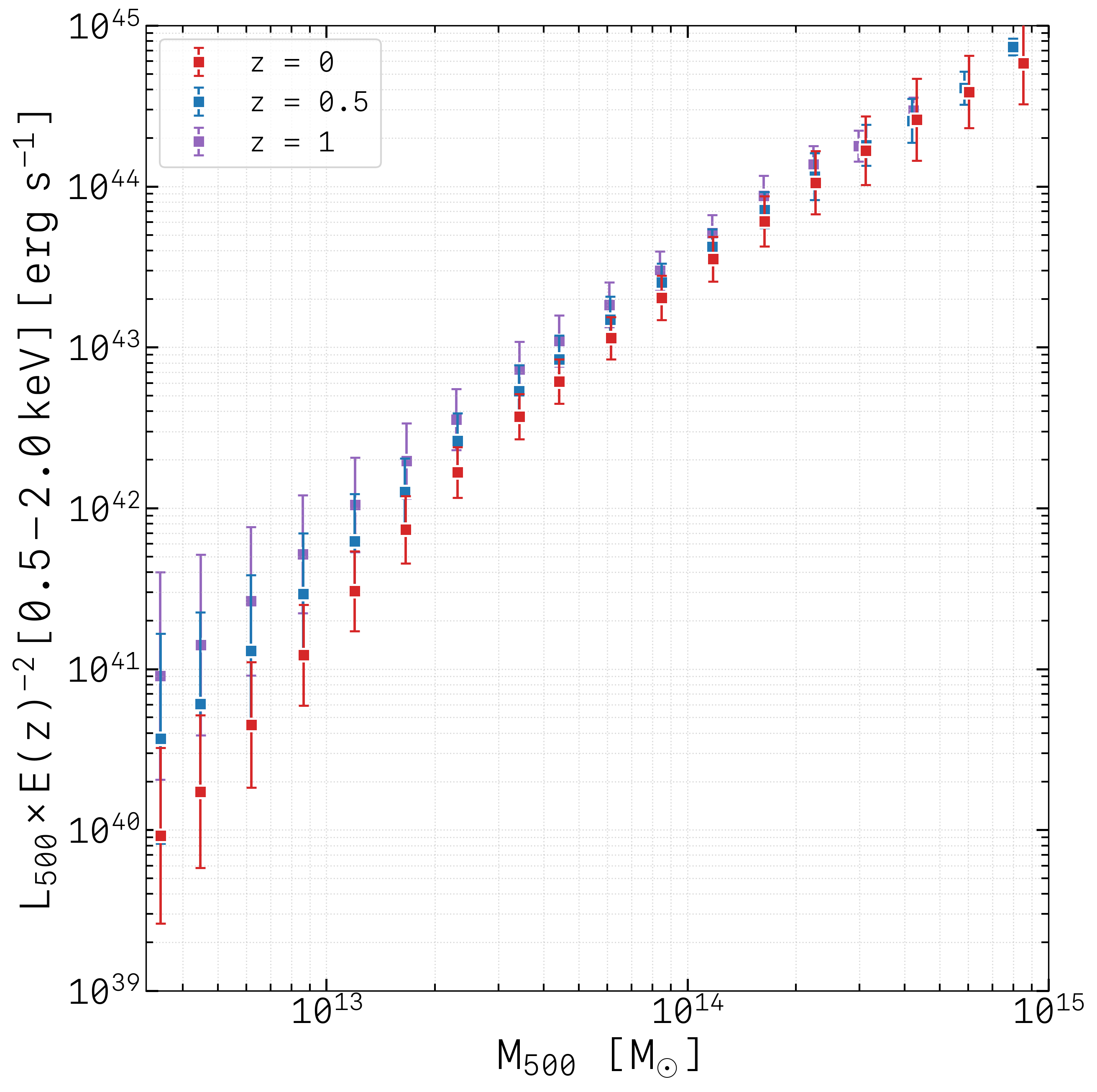}
\caption{Evolution of the soft-band ($0.5$--$2.0\,\mathrm{keV}$) X-ray luminosity--mass scaling relation $L_{500}$--$M_{500}$ at redshifts $z = 0$, $0.5$, and $1$, derived from the TNG300 and MTNG simulations. At the high-mass end ($M_{500} \gtrsim 10^{14}\,\mathrm{M}_\odot$), the three relations overlap, consistent with the self-similar expectation for systems whose thermodynamic state is governed by gravitational processes. At lower masses, a systematic offset toward higher luminosities is observed at earlier epochs, reflecting the progressive depletion of hot gas from shallower potential wells due to AGN feedback over cosmic time. The increasing scatter and steepening of the slope at group scales signal the growing dominance of non-gravitational physics in this regime.}
\label{Figure:Redshift}
\end{figure}

While our analysis has thus far focused on the ICM properties at $z=0$, characterizing the redshift evolution of the $L_X - M_{\rm 500}$ scaling relation is essential for breaking the degeneracy between gravitational heating and non-gravitational feedback processes. In Figure \ref{Figure:Redshift}, we present the evolution of the X-ray luminosity-mass relation for galaxy groups and clusters across redshifts $z = 0$, $0.5$, and $1$. At the high-mass end ($M_{\rm 500} \geq 10^{13.67} \,\text{M}_{\odot} )$, the tight overlap of the relations confirms that massive clusters behave largely as closed boxes, where the deep gravitational potential retains baryons effectively against feedback mechanisms. These systems evolve in agreement with the self-similar expectation, indicating that their thermodynamic state is governed primarily by gravitational shock heating during hierarchical collapse~\citep{Voit2005}.

Conversely, the low-mass regime reveals a distinctive evolutionary trend: at fixed mass, high-redshift groups appear systematically more luminous than their local counterparts. This vertical offset signifies a strong departure from self-similarity, driven by non-gravitational physics such as AGN feedback and radiative cooling. Simulations indicate that this evolution is driven by the progressive evacuation of hot gas from the shallower potential wells of groups; as feedback mechanisms (particularly kinetic AGN activity) operate over cosmic time, they deplete the baryon fraction and suppress X-ray luminosity by $z=0$ compared to earlier epochs~\citep{Puchwein2008, LeBrun2014,  Ragagnin2022, Robson2023, Marini2025}. Consequently, the steepening of the slope and the increase in scatter at lower masses reflect the growing dominance of baryonic physics over gravity in the group regime.

This characteristic break in the $L_X - M_{\rm 500}$ can be proposed as a physically motivated boundary between the galaxy group and cluster regimes with a pivot mass in the range $M_{\rm 500}\approx 10^{14}\, \text{M}_{\odot}$. The transition marks the threshold where AGN feedback energy becomes comparable to the halo's gravitational binding energy. Below this mass, feedback mechanisms expel gas from the shallower potential wells of groups, decreasing X-ray luminosity and driving the system away from self-similarity. Conversely, massive clusters retain their baryon content, behaving as closed boxes that follow the self-similar predictions. This analysis constitutes the first study of the redshift evolution of the soft-band $L_X$--$M_{500}$ scaling relation within the TNG physical model using both TNG300 and MTNG, providing a direct prediction for the evolution of the group and cluster population accessible to current eROSITA surveys.

 \section{Conclusions and Discussion}  
 \label{Sec:Conclusions}

This paper presents an optimised pipeline to generate X-ray mocks for galaxy clusters in the TNG300 and MTNG simulations. This pipeline allows us to compute the luminosity distribution of the galaxy clusters using the APEC code developed to estimate the emission spectrum. This code can process large volumes of data in a relatively short time ($\sim$ 4 hours for a TNG300 or a MTNG snapshot). The combination of TNG300 and MTNG allowed us to compute the $L_{X, \rm 500} - M_{\rm 500}$ scaling relation over a wide mass range, from $10^{12.5}\, \text{M}_{\odot}$ to $10^{15.5}\, \text{M}_{\odot}$ and to have good statistics for the whole mass range. We validate the pipeline used to create the X-ray mocks by comparing the simulated scaling relation with the observational data from multiple observations, especially in the range of high-mass clusters, $M_{\rm 500}> 10^{13.67} \,\text{M}_{\odot}$ \citep{Eckmiller2011, Mantz2016a, Lovisari2015}. The low-mass cluster samples are limited. We hope that new X-ray surveys, such as eROSITA, can improve the coverage of this mass range, helping us to understand the physical processes in the presence of a broken power law. 

The slopes derived from the simulated scaling relations were compared with those predicted by self-similarity theory, described in Section~\ref{Sec:Introduction}.  We find that, for high-mass clusters, the slopes are relatively close to the theoretical predictions. However, for low-mass clusters, the simulated slopes are steeper than those predicted by self-similarity. Non-gravitational processes, such as galactic winds, supernova feedback, cosmic rays, and active galactic nuclei (AGN), are expected to have a more significant impact on these smaller structures due to their shallower gravitational potential, increasing the intrinsic scatter in the luminosity and altering the global properties of galaxy groups~\citep{Eckmiller2011}. For example, \citet{Sijacki2007}, \citet{McCarthy2010}, and \citet{Short2010} demonstrated that incorporating AGN feedback into cosmological simulations disrupts self-similarity, lowering the luminosities.

In this paper, we also disentangle the effects of several assumptions commonly adopted in the analysis of X-ray observations. Our results demonstrate that, even under idealized conditions, the spectroscopic procedure introduces systematic and mass-dependent biases in both the recovered halo mass and \mbox{X-ray} luminosity \citep{Piffaretti2008, Barnes2021, Braspenning2025}. In particular, the recovered luminosity is lower than the simulation-based value by approximately $18\%$ at the high-mass end and by up to $33\%$ in lower-mass systems, while the recovered mass is underestimated by roughly $15\%$ in the low-mass regime and up to $21\%$ toward higher masses. These trends indicate that the bias introduced by the X-ray analysis is not uniform, but depends on the physical properties of the system and on the assumptions entering the spectroscopic reconstruction~\citep{Barnes2021, Braspenning2025}. Low-mass halos are more sensitive to line emission, multiphase gas structure, and reduced photon statistics, which amplifies the impact of fitting each radial bin with a single-temperature, single-metallicity model~\citep{Piffaretti2008, Pearce2019}. At the same time, the mass estimate remains sensitive to the recovered temperature structure and to departures from hydrostatic equilibrium~\citep{Piffaretti2008, Braspenning2025}. Because the spectroscopic pipeline shifts both mass and luminosity simultaneously, part of the bias is absorbed along the intrinsic $M_{500}$--$L_{500}$ relation rather than perpendicular to it~\citep{Pratt2009, Lovisari2020}, leaving the global normalization and slope only weakly affected even though the individual estimates of $M_{500}$ and $L_{500}$ remain biased with respect to their simulation-based values. This result highlights that agreement at the level of the scaling relation does not necessarily imply unbiased recovery of the underlying physical properties, but instead reflects the coupled way in which the observational procedure modifies both variables~\citep{Barnes2021}.

More importantly, we demonstrate that the departures from self-similarity observed in the galaxy group regime are not merely artifacts of the mass reconstruction or X-ray luminosity measurement procedures. Instead, these deviations are primarily driven by non-gravitational baryonic processes, which exert a larger impact on the shallower potential wells of low-mass systems. While the fundamental assumptions of hydrostatic equilibrium and spherical symmetry do introduce systematic biases in total mass estimates and recovered luminosities, our results indicate that these measurement-induced errors remain relatively uniform across the sampled mass range. Consequently, the characteristic scaling breaks and increased scatter observed in the group regime must be attributed to physical phenomena, such as the efficient ejection of gas from the centers of low-mass halos due to feedback mechanisms.

While our mock X-ray analysis incorporates realistic measurement biases, it is important to note that we focused on individual clusters in isolation and did not consider projection effects like contamination due to background or foreground emission.  \citet{Shreeram2025} investigated such projection effects using synthetic eROSITA light-cones, finding that they can slightly boost the measured $L_{X}$ or alter the apparent morphology of groups and clusters, though X-ray selection is still far less susceptible to projection than optical or lensing selection. In the context of our results, projection effects could introduce additional scatter or slight biases in the observed scaling relations. For example, a cluster behind a filament or group might appear more luminous, due to added line-of-sight emission,  than it truly is. However, the density-squared dependence of X-ray emission means truly massive halos dominate the X-ray signal, and the probability of major projection misidentification is lower compared to other wavelengths such as optical cluster catalogues. Thus, we expect our conclusions about mass and luminosity biases to hold in general, with projection contributing as a second-order effect. Nonetheless, as new X-ray surveys with higher sensitivity come online, a  careful background modelling and cross-matching will remain crucial to disentangle any projection-enhanced signals, as done in \citealt{Marini2024}.

\begin{acknowledgements}
FA and AS acknowledge support by the Italian Research Center on High Performance Computing, Big Data and Quantum Computing (ICSC), project funded by European Union - NextGenerationEU - and National Recovery and Resilience Plan (NRRP) - Mission 4 Component 2, within the activities of Spoke 3, Astrophysics and Cosmos Observations; by the PRIN 2022 PNRR project (202259YAF) “Space-based cosmology with Euclid: the role of High-Performance Computing.”; by the PRIN 2022 (20225E4SY5) “From ProtoClusters to Clusters in one Gyr”; by the INAF Astrofisica Fondamentale Large Grant 2023 “Witnessing the Birth of the Most Massive Structures of the Universe”. FA and AS acknowledge ISCRA for awarding this project access to the LEONARDO supercomputer, owned by the EuroHPC Joint Undertaking, hosted by CINECA (Italy) . FAG acknowledges support from the ANID BASAL project FB210003, from the ANID FONDECYT Regular grant 1251493 and from the HORIZON-MSCA-2021-SE-01 Research and Innovation Programme under the Marie SklodowskaCurie grant agreement number 101086388. TC is supported by the Agenzia Spaziale Italiana (ASI) under – Euclid-FASE D Attivita’ scientifica per la missione – Accordo attuativo ASI-INAF n. 2018-23-HH.0, by the National Recovery and Resilience Plan (NRRP), Mission 4, Component 2, Investment 1.1, Call for tender No. 1409 published on 14.9.2022 by the Italian Ministry of University and Research (MUR), funded by the European Union – NextGenerationEU– Project Title “Space-based cosmology with Euclid: the role of High-Performance Computing” – CUP J53D23019100001 – Grant Assignment Decree No. 962 adopted on 30/06/2023 by the Italian Ministry of University and Research (MUR); by the Italian Research Center on High-Performance Computing Big Data and Quantum Computing (ICSC) a project funded by European Union – NextGenerationEU – and National Recovery and Resilience Plan (NRRP) – Mission 4 Component 2. FA, TC, and AS acknowledge the support by the INFN INDARK PD51 grant. F.A. acknowledges financial support from ANID through the FONDECYT Postdoctoral Fellowship, Grant No. 3260589.
\end{acknowledgements}

%
\bibliographystyle{aa} 
\bibliography{references} 
%
\begin{appendix}
\section{Scaling relations without excluding the centres of galaxy clusters}
\label{Sec:Mass_baias_todo}
\begin{figure}
\centering
\includegraphics[width=\linewidth]{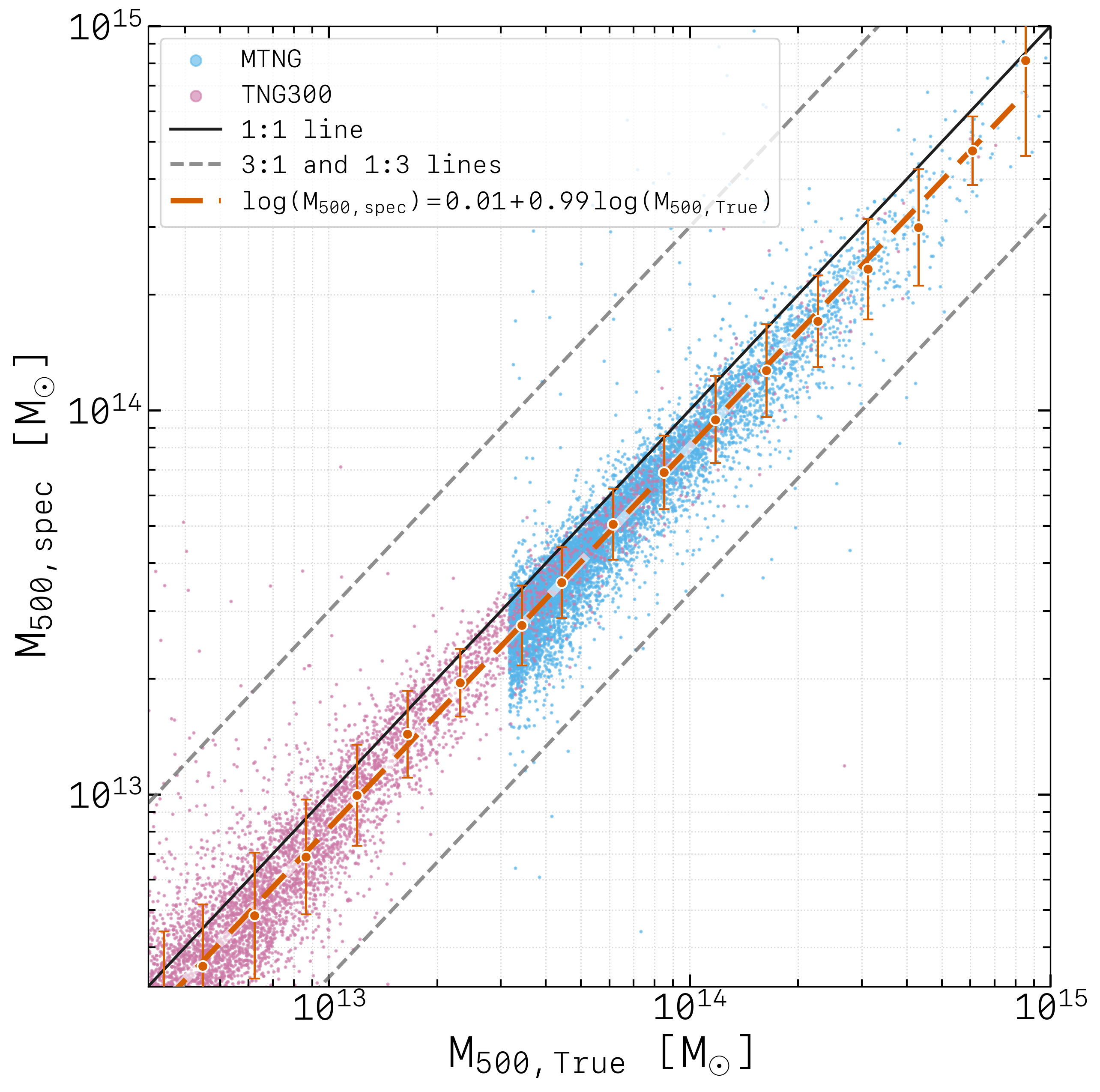}

\caption{Comparison between the true halo mass $M_{500,\mathrm{true}}$ and the spectroscopic mass $M_{500,\mathrm{spec}}$ derived from the mock X-ray pipeline when the full cluster volume within $R_{500}$ is used, without excluding the central region. In this case, the mass offset is approximately $20\%$ and is nearly uniform across the full mass range, in contrast to the mass-dependent bias found when the central regions are excluded (Section~\ref{Sec:Mass_baias}).}
\label{Fig:spec_true_todo}
\end{figure}
\begin{figure}
\centering
\includegraphics[width=\linewidth]{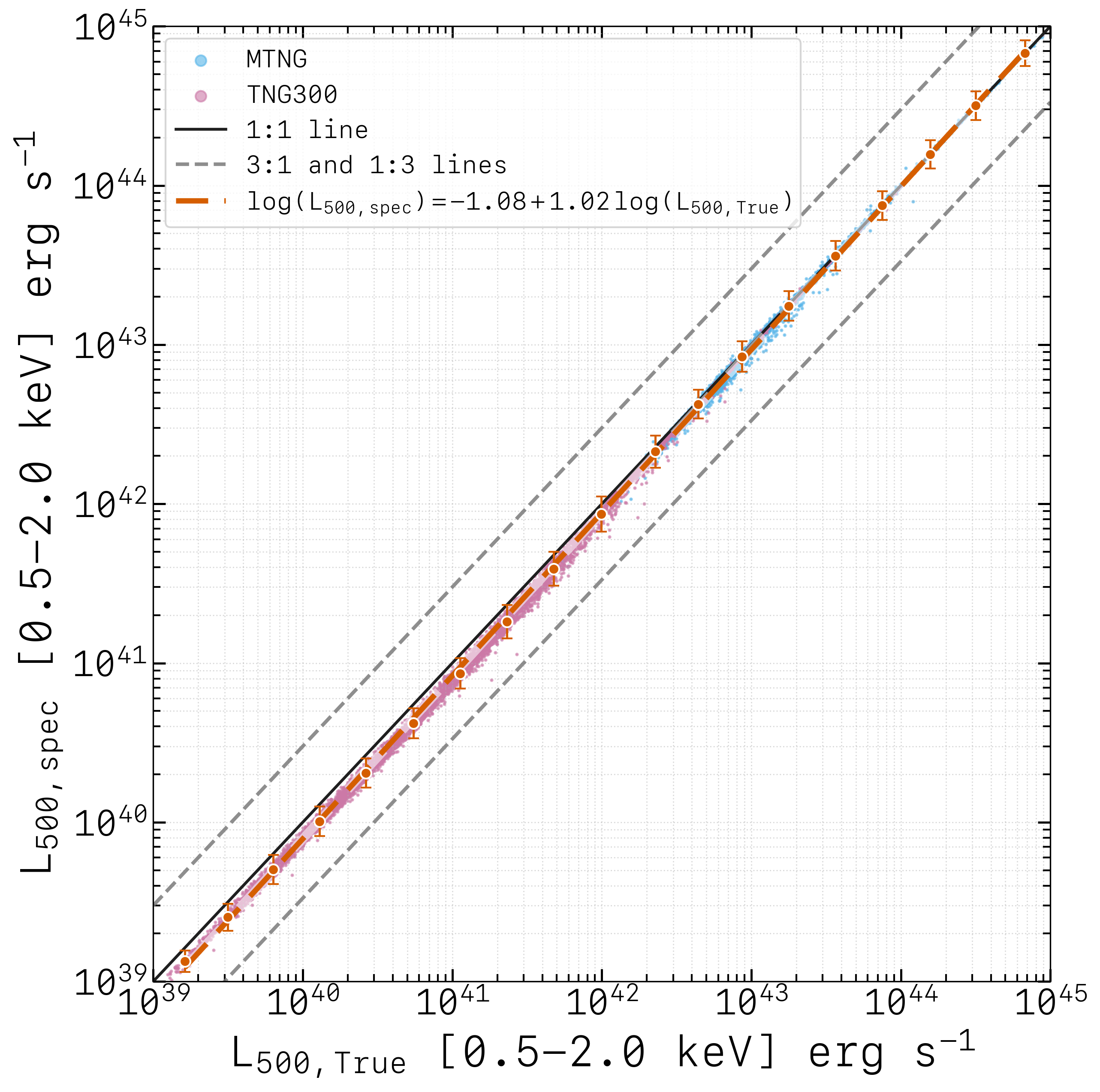}
\caption{Comparison between the true X-ray luminosity $L_{500,\mathrm{true}}$ and the spectroscopic luminosity $L_{500,\mathrm{spec}}$ recovered from the mock X-ray pipeline when the full cluster volume within $R_{500}$ is used, without excluding 
the central region. The one-to-one relation is shown by the dashed line. The luminosity offset is reduced compared to the case with central region exclusion: the discrepancy is approximately $2\%$ for high-mass clusters and approximately $15\%$ for low-mass systems. The improved recovery arises because the spectral fit includes the emissivity contribution from the dense central gas.}
\label{Fig:Xrays_true_spec_todo}
\end{figure}

In this Appendix, we present the derivation of the scaling relations similar to the analysis presented in Section~\ref{Sec:Mass_baias}, but including also the central regions of the galaxy clusters. As shown in Figures~\ref{Fig:spec_true_todo}, and \ref{Fig:Xrays_true_spec_todo}, the offsets of the mass and luminosity spectroscopically derived mass are closer to the true values compared to the case when we exclude the central regions, and the scatter is reduced. In the case of the mass, the spectroscopically derived values underestimate the true mass by around $20\%$, but this offset is almost the same for the full mass range. 

In Figure~\ref{Fig:Xrays_true_spec_todo}, we compare the X-ray luminosity $L_{\rm 500}$ measured directly from the simulation (summing hot gas particle emission) against the $L_{\rm 500}$ obtained from spectral fitting of the mock X-ray maps. The derived luminosity has an offset of 15\% for low mass clusters and a 2\% offset for the high mass clusters.  This smaller bias arises because, while the mass and temperature determinations are affected by biases as discussed above, the luminosity is an integrated quantity dominated by emission from dense, cool gas, mainly concentrated in the central regions of galaxy clusters.  That density is effectively accounted for in the spectral fit normalization when we use the data corresponding to the whole cluster. 

\begin{figure}
\centering
\includegraphics[width=\linewidth]{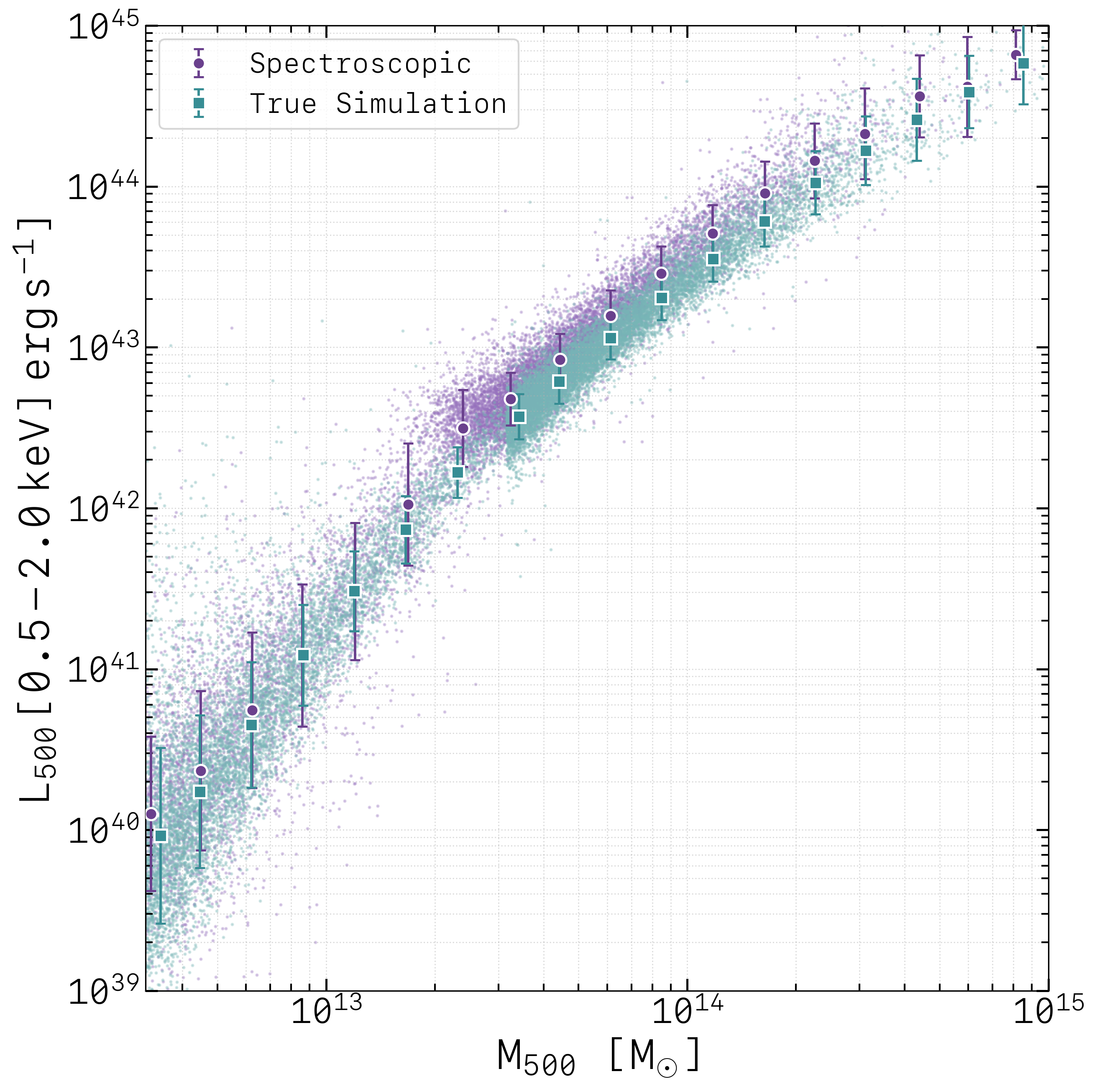}
\caption{Comparison of the $M_{500}$--$L_{500}$ scaling relation derived from two definitions of the underlying quantities, using the full cluster volume within $R_{500}$ without central region exclusion. 
Green points show the simulation-based relation $(M_{500,\mathrm{true}},\,L_{500,\mathrm{true}})$; purple points show the spectroscopic relation $(M_{500,\mathrm{spec}},\,L_{500,\mathrm{spec}})$ from the mock X-ray pipeline. Unlike the case with central region exclusion (Figure~\ref{Fig:sim-spec}), a systematic offset in normalisation is present between the two relations, particularly in the cluster regime, arising from the approximately uniform $20\%$ mass underestimation. The slope of the $M_{500}$--$L_{500}$ relation remains consistent between the two cases, with values of $\alpha = 1.2$ at the 
high-mass end and $\alpha = 2.3$ in the group regime, in agreement with the results derived directly from the simulation outputs in Section~\ref{sec:Scaling}.}
\label{Fig:sim-spec_todo}
\end{figure}

Finally, in Figure~\ref{Fig:sim-spec_todo}, we present the halo mass–luminosity ($M_{\rm 500}$–$L_{\rm 500}$) scaling relations derived in the two cases: the ``true'' $(M_{\rm 500, SIM}  -  L_{\rm 500, SIM})$ values in green, versus ``observed'' ($M_{\rm 500, spec}  - L_{\rm 500, spec}$) values in purple; the overall slope of the scaling relations remain very similar, both roughly follow a broken power law with a steepening at group scales,  but there is a clear normalization offset. At a given true mass, the X-ray-derived mass is smaller, so one could assign that cluster to a lower mass bin observationally. Consequently, when binned by the X-ray mass, the average luminosity appears higher than it would at the true mass. This is why in Figure~\ref{Fig:sim-spec_todo} the purple points are shifted toward higher $L_{\rm 500}$ at fixed $M_{\rm 500}$ compared to the intrinsic relation for cluster regime, while for the group regime the scaling relations remains almost unchanged. We find that the spectroscopic relation is offset to higher luminosity by roughly the amount expected from a 20\% mass bias. Despite these differences in normalization and scatter, the slope of the $M - L$ relation remains consistent between true and mock observations. This is a key result: the self-similar scaling expected for massive clusters (e.g. $L \propto M^{1}$ in the soft band) is preserved at the high-mass end, where we measure a slope of $1.2$, whereas in the group-mass regime the relation steepens to a slope of $2.3$. Both values are in very good agreement with those obtained when the luminosities and masses are computed directly from the hydrodynamical simulations derived in Section~\ref{sec:Scaling}.

Thus, these results show that X-ray-derived masses are systematically biased low, typically by a roughly constant fraction of 20\% when compared to the true gravitating mass. However, the shape and slopes of the scaling relations remain largely governed by gravity, as predicted by the self-similar model of structure formation~\citep{Rasia2006, Lau2009, Biffi2016, Angelinelli2020, Ansarifard2020}.  Because \mbox{X-ray} emissivity scales with the square of the gas density, the \mbox{X-ray} luminosity of clusters drops more drastically than their temperature during feedback events. This results in a steeper $L_{X}$–$M_{\rm 500}$ slope for low-mass systems, whereas the $T$–$M_{\rm 500}$ relation remains remarkably close to the self-similar prediction across both group and cluster mass scales~\citep{Toptun2025}.

\section{Element abundance approximation}
\label{Ap:Metals_metallicity}

Some cosmological simulations, such as IllustrisTNG, explicitly trace the abundances of individual chemical elements, whereas others, such as MillenniumTNG, only track the total metallicity (this was done here to realize necessary memory savings). Accurate metal abundance information is particularly relevant for gas particles with temperatures $T \leq 10^{6}\,\mathrm{K}$, where the X-ray emission is dominated by line emission. In this regime, the primary contributors are the metal lines from heavy elements (e.g., Fe, O, Ne), making reliable metal tracking especially critical in the cool cores of massive clusters and in low-mass halos, where the gas temperature remains relatively low \citep{Bohringer2010, Mulchaey2000}. 

To assess the impact of this limitation, we compared the $L_X$-$M_{\rm 500}$ scaling relation obtained using explicitly traced elemental abundances with that derived from an approximation based solely on the total metallicity. Across all mass bins, we find that the X-ray luminosities derived using the metallicity-based approximation closely follow those computed using individual element abundances. This agreement validates our approximation method and supports its use as a robust approach for modeling X-ray emission in simulations where individual metal abundances are unavailable.
\begin{figure}[h]
\centering
\includegraphics[width=\linewidth, trim={0.0cm, 0.4cm, 1.8cm, 1.1cm}, clip]{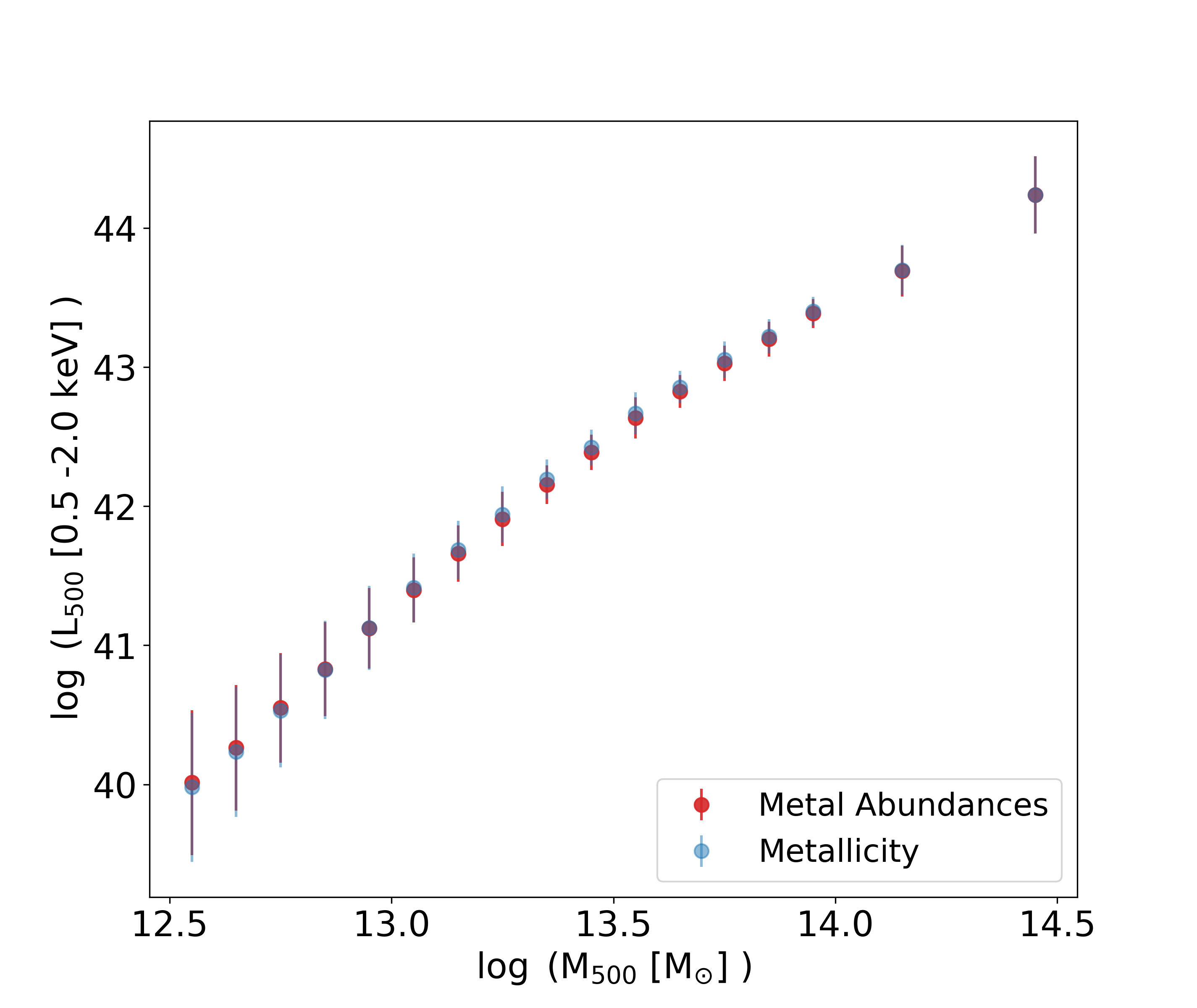}
\caption{Comparison of the $L_{500}$--$M_{500}$ scaling relation obtained using two different approaches to model the ICM chemical composition in TNG300. Blue points show the relation derived using the individual elemental abundances explicitly tracked by the simulation, while red points show the relation obtained using the 
metallicity-based approximation described in Section~\ref{sec:Mocks}. The close agreement between the two approaches across the full mass range validates the metallicity-based prescription as a reliable method for generating X-ray emission in 
simulations that do not track individual chemical elements, with a mean luminosity deviation below $2\%$.}
\label{Metals_metallicity}
\end{figure}

\section{Derivation of the total mass}
\label{Ap:profiles}
One of the most important quantity commonly derived from \mbox{X-ray} observations are  mass estimates. The key assumption for this derivation is that the hot gas is inside a spherically symmetric potential well and is close to hydrostatic equilibrium. The gas profile can then be computed from 
\begin{equation}
    M_{\rm gas}(<r) =4\pi \frac{A}{Z} m_p \int_{0}^r n_e(r)r^2\, {\rm d}r,
\end{equation}
where $m_p$ is the mass of protons, $A \approx 1.4$ is the mean nuclear
mass number, and $Z \approx 1.2$ is the charge number of the ICM with 0.3 solar abundance. To obtain the total mass, we will assume that galaxy clusters are in hydrostatic equilibrium which is not true for all clusters. This assumption means that the pressure gradient is in balance with the gravity. Then the total mass of a galaxy cluster can be derived using: 
\begin{equation}
    M(<r)=-\frac{k_B T(r) r}{G\mu m_p}\left(\frac{d \ln \rho_g (r)}{d\ln r} + \frac{d\ln T (r)}{d \ln r }\right),
    \label{Hydrostatic_equilibrium_equation}
\end{equation}
where $\mu $ is the mean molecular weight and $T(r)$ and $\rho_g(r)$ are the temperature and density profiles obtained fitting the X-ray spectrum. 

To fit the temperature profile, we use the McDonald model, given in the following way: 
\begin{equation}
T(r)= T_0 \frac{(r/r_{\rm cool})^\alpha +T_{\rm min}/T_0}{(r/r_{\rm cool})^{a_{\rm cool}}+1} \cdot \frac{(r/r_t)^{-a}}{[1+(r/r_t)^b]^{c/b}} .
\label{Eq:temp_profile}
\end{equation}
Then the temperature profile has eight free parameters,
and the density profile can be fitted using the $\beta$-model \citep{Vikhlinin2006} given by:
\begin{equation}
    n_pn_e= n_0^2 \frac{(r/r_c)^\alpha}{(1+r^2/r_c^2)^{3\beta-\alpha/2}} \cdot \frac{1}{(1+r^\gamma/r_s^\gamma)^{\epsilon/\gamma}} ,
\label{Eq:density}
\end{equation}
where we fixed $\gamma = 3$. Both the temperature and density parameters have enough freedom to fit most of the profiles present in the halos selected for this work. The profiles for which no best-fit parameters could be found are ignored.

\end{appendix}

\end{document}